\documentclass[aps,prc,twocolumn,groupedaddress,floatfix]{revtex4-2}

\usepackage{amsmath}
\usepackage{amssymb}
\usepackage{slashed}
\usepackage{float}
\usepackage{graphicx}
\usepackage{xcolor}
\usepackage{subfigure}
\usepackage{url}

\definecolor{mypink}{rgb}{0.858, 0.188, 0.478}

\usepackage{ulem}
\usepackage{lineno}
\usepackage{verbatim}

\begin{document}

%\preprint{}

\title{Second Class Currents, Axial Mass and Nuclear Effects in Hyperon Production}


\author{Christopher Thorpe}
%\thanks{}
\affiliation{Department of Physics, Lancaster University, LA1 4YB, Lancaster, United Kingdom\looseness=-1}
\author{Jarosław Nowak}
\affiliation{Department of Physics, Lancaster University, LA1 4YB, Lancaster, United Kingdom\looseness=-1}
\author{Kajetan Niewczas}
\affiliation{Institute of Theoretical Physics, University of Wrocław, pl. M. Borna 9, 50-204, Wrocław, Poland \looseness=-1}
\affiliation{Department of Physics and Astronomy, Ghent University, Proeftuinstraat 86, B-9000 Ghent, Belgium.}
\author{Jan T. Sobczyk}
\affiliation{Institute of Theoretical Physics, University of Wrocław, pl. M. Borna 9, 50-204, Wrocław, Poland \looseness=-1}
\author{Cezary Juszczak}\looseness=-1
\affiliation{Institute of Theoretical Physics, University of Wrocław, pl. M. Borna 9, 50-204, Wrocław, Poland \looseness=-1}


\date{\today}

\begin{abstract}
We study the properties of the Cabibbo suppressed quasielastic production of $\Lambda$ and $\Sigma$ hyperons in antineutrino interactions with nuclei, including the effects of modified form factor axial mass, the second class current and SU(3) flavour violations. The hyperon and nucleon are subjected to the nuclear potential and the outgoing hyperon can undergo final state interactions. The hyperon potential has a significant effect on their production through absorption. We predict a significant enhancement of $\Lambda$ production compared with other hyperon production channels through $\Sigma \to \Lambda$ conversions. We produce predictions for several experiments by combining realistic neutrino energy distributions with suitable nuclear targets. 
\end{abstract}

%\keywords{}

\maketitle



\section{Introduction}

In this paper we discuss the weak production of the four hyperons $\Lambda,\Sigma^{\pm,0}$ belonging to the $J^P=1/2^+$ baryon octet, three of which are produced directly in interactions between antineutrinos and nucleons: 

\begin{align}
\bar{\nu}_l + p &\to \Lambda + l^+ \\
\bar{\nu}_l + p &\to \Sigma^0 + l^+  \\
\bar{\nu}_l +n &\to \Sigma^- + l^+
\end{align}

The $\Sigma^+$ is produced by reinteraction of the initial hyperon inside the nucleus. A generic hyperon production process is notated as $\bar{\nu}_l + N \to Y + l^+$.
Fewer than 100 events of this type have been observed so far \cite{Ammosov:1986jn,Eichten:1972bb,Erriquez1,Erriquez2,Brunner:1989kw,Fanourakis:1980} and the parameters of the theoretical models are not well constrained by data as a result. Instead, symmetry arguments such as the SU(3) quark flavour model are invoked to get the parameters from the better understood charged current quasielastic (CCQE) process:

\begin{equation}
\bar{\nu}_l + n \to p + l^+ .
\end{equation}

Future experiments such as the Short Baseline Near Detector (SBND) and the Deep Underground Neutrino Experiment (DUNE) will be able to observe far larger numbers of hyperon production events, with 12,500 predicted for SBND over 3 years of running~\cite{Brailsford:2017rxe}. With these statistics it will be possible to test various hypotheses in hadronic physics, such as the validity of the underlying symmetries in hadron production. Hyperons are of similar mass to nucleons however they are not subjected to the same nuclear effects, making them unique nuclear probes.

In this work we explore the effects of including SU(3) symmetry breaking corrections, G parity and time reversal violations (second class currents). The absence of second class currents was first postulated by Weinberg~\cite{Weinberg:1958ut} and implies a connection between isospin and weak interactions. The effects of the nuclear potential are studied through the inclusion of nucleon-nucleus and hyperon-nucleus potentials and an intranuclear cascade. 

We implemented the model in the NuWro Monte Carlo generator \cite{Nowak:2006cp} as a new process which represents an advancement to the simulation of hyperon production compared to other generators that use a model from Pais \cite{Pais71}. The NuWro generator allows for detailed analysis of how the nuclear effects and final state interactions alter the kinematics of the final state hyperons. 

This work is organised as follows: Section \ref{Section2} details the cross section calculation, including the second class current and symmetry breaking corrections. Section \ref{Section3} describes the intranuclear cascade used to model reinteractions between the hyperon and nuclear remnant. Comparisons between data and predicted cross sections for hyperon production from free nucleons are presented in section \ref{Section4}. Nuclear effects are studied in section \ref{Section5}, such as the influence of the hyperon potential on differential cross sections. Realistic fluxes are combined with nuclear targets in section \ref{Section6} to explore different observable quantities and their sensitivity to features of the hyperon production model.

\section{Theory}
\label{Section2}
\subsection{Cross Section Model}

The general expression for the  differential cross section for hyperon production is~\cite{Fatima:2018tzs}:

\begin{equation}
\frac{d\sigma}{d Q^2} = \frac{G_F^2 \sin^2{\theta_c}}{8 \pi E_{\nu}^2 M_{N}^2} \mathcal{L^{\mu\nu}}\mathcal{H_{\mu\nu}}   
\end{equation}

The reaction takes place on a bound nucleon. We adopt the following notation: $E_{\nu}$ is the energy of the neutrino in the nucleon rest frame and $M_N$ is the mass of the target nucleon. The interaction involves a bound nucleon and a hyperon interacting with a potential and this is accounted for in how the kinematics of the particles are calculated. This is performed using effective masses, $M_X^*$, related to free mass $M_X$, the momentum $\textbf{p}_X$ and potential energy $V$ by:

\begin{equation}
M_{X}^* = \sqrt{\left(\sqrt{M_{X}^2+|\textbf{p}_{X}|^2}-V\right)^2-|\textbf{p}_{X}|^2 } 
\end{equation}

For nucleons this potential is the nuclear binding energy and for hyperons the potential described in section IID. The effective masses are used in place of the real masses when determining the kinematics of outgoing particles. Leptons are not subjected to any nuclear effects in this work.

Working in the center of mass frame, we denote the energy of the neutrino and 3 momenta of the outgoing charged lepton as $E_{\nu}$ and $\textbf{k}_l$ respectively. When the scattering angle in this frame is $\theta$ the squared 4 momentum transfer $Q^2=-q^2=(k-k_l)^2$ is:

\begin{equation}
Q^2 = 2E_{\nu}\sqrt{m_l^2+|\textbf{k}_l|^2} - m_l^2 - 2E_{\nu}|\textbf{k}_l|\cos{\theta}
\end{equation}

$\mathcal{L}^{\mu\nu}$ and $\mathcal{H}_{\mu\nu}$ are the leptonic and hadronic tensors, constructed by summing/averaging the hadronic and leptonic currents over the initial/final state particle spins:

\begin{align}
\mathcal{H}^{\mu\nu} &= \frac{1}{2} \textrm{Tr} \left[\Gamma^{\mu}(\slashed{p}_{N}+M_{N})\tilde{\Gamma}^{\nu}(\slashed{p}_Y+M_{Y}) \right] \\
\mathcal{L}^{\mu\nu} &=  \textrm{Tr}\left[ \gamma^{\mu}(1+\gamma_5)\slashed{k}\gamma^{\nu}(1+\gamma_5)(\slashed{k_l}-m_l) \right]
\end{align}

The leptonic current is described by V-A theory~\cite{Lee:1957qr}. To account for the composite structure of the nucleon the hadronic current involves six Dirac operators, the effective couplings of which are described by a set of dimensionless form factors:

\begin{align}
\Gamma^{\mu} &=V^{\mu} - A^{\mu} \\
 V^{\mu} &= f_1(Q^2)\gamma^{\mu} + if_2(Q^2)\sigma^{\mu\nu}\frac{q_{\nu}}{M} + f_3(Q^2)\frac{q^{\mu}}{M}   \\
 A^{\mu} &= \left[ g_1(Q^2)\gamma^{\mu} - ig_2(Q^2)\sigma^{\mu\nu}\frac{q_{\nu}}{M} + g_3(Q^2)\frac{q^{\mu}}{M}\right]\gamma_5
\end{align}

Where $M=M_{N} + M_{Y}$, the sum of the masses of the nucleon and hyperon.

\subsection{Form Factors}

The approach for obtaining the form factors is the one described in \cite{Fatima:2018tzs}, briefly summarised:

\begin{enumerate}
    \item Time reversal invariance implies $f_i(Q^2)$ and $g_i(Q^2)$ are real.
    \item Conservation of vector current (CVC) requires $f_3(Q^2)=0$.
    \item G parity invariance leads to $f_3(Q^2)=0$ and $g_2(Q^2)=0$. This pair of form factors describe the coupling of the second class current (SCC).
    \item The remaining form factors are obtained in terms of their counterparts in strangeness conserving quasielastic scattering using the SU(3) symmetry and partially conserved axial current (PCAC).
\end{enumerate}

The form factors $f_{1,2}^{p,n}$ are expressed in terms of the nucleon electric and magnetic form factors, $G_{E,M}^{p,n}$ for which the BBBA05 parameterisation \cite{Bradford1}  has been used. The hyperon form factors themselves are listed in table \ref{Table1}. %% Should we compare dipole and BBBA ?

\begin{align}
f_1^{p,n}(Q^2) &= \frac{1}{1+\tau}\left[G_E^{p,n}(Q^2) - \frac{1}{1+\tau}G_M^{p,n}(Q^2) \right] \\
f_2^{p,n}(Q^2) &= \frac{1}{1+\tau}\left[G_M^{p,n}(Q^2) - G_E^{p,n}(Q^2) \right] 
\end{align}

The pseudoscalar form factor $g_3$ is derived using PCAC by Nambu \cite{Nambu:1960xd} in terms of the axial form factor:

\begin{align}
g_3^{NY} = \frac{(M_N + M_Y)^2}{2(m_K^2 + Q^2)}g_1^{NY}
\end{align}

$m_K = 0.498$ GeV is the $K^0$ mass and $M_Y$ is the hyperon mass.

\begin{table}[H]
\centering
\begin{tabular}{c|ccc}
\hline
     & $f_1(Q^2)$ & $f_2(Q^2)$ & $g_1(Q^2)$ \\
     \hline 
  $p \to \Lambda $   & $-\sqrt{\frac{3}{2}}f_1^p$ &  $-\sqrt{\frac{3}{2}}f_2^p$ & $-\frac{1}{\sqrt{6}}\left(1+2x \right)g_A$  \\
  \hline
  $p \to \Sigma^0$ & $- \frac{1}{\sqrt{2}}\left( f_{1}^{p} + 2 f_{1}^{n} \right)$ & $- \frac{1}{\sqrt{2}}\left( f_{2}^{p} + 2 f_{2}^{n} \right)$ & $\frac{1}{\sqrt{2}}\left(1-2x \right)g_A$ \\
  \hline
  $n \to \Sigma^-$ & $- \left( f_{1}^{p} + 2 f_{1}^{n} \right)$ & $- \left( f_{2}^{p} + 2 f_{2}^{n} \right)$ & $\left(1-2x \right)g_A$ \\
  \hline
\end{tabular}

\caption{List of form factors. The quasielastic axial form factor $g_A$ satisfies $g_A(0)=1.2673$ described by a dipole of axial mass $M_A$ and $x=0.365$. Unless otherwise specified we use $M_A = 1$ GeV.}
\label{Table1}
\end{table}

Comparison of the form factors for $p\to\Sigma^0$ and $n\to\Sigma^-$ results in the following relation in the case of unbroken SU(3) symmetry:

\begin{align}
\frac{d\sigma}{dQ^2}(\bar{\nu}_{\mu}p\to\mu^+\Sigma^0) \approx \frac{1}{2}\frac{d\sigma}{dQ^2}(\bar{\nu}_{\mu}n\to\mu^+\Sigma^-)    
\end{align}

An equality if $M_n=M_p$ and $M_{\Sigma^0}=M_{\Sigma^-}$.

\subsection{Second Class Current and Symmetry Breaking Effects}

The vector and axial currents of the standard model transform under G parity the following way:

\begin{align}
GV^{\mu}G^{-1} = V^{\mu} \\
GA^{\mu}G^{-1} = - A^{\mu}
\end{align}

Current operators that transform according to these rules are the so called first class currents. In equations 9-11, the operators associated with $f_1$, $f_2$, $g_1$ and $g_3$ transform this way, whilst those associated with $f_3$ and $g_2$ do not, these are the second class currents. 

The scalar interaction controlled by $f_3$ is also forbidden by CVC and is not considered in this work. Here we consider the reintroduction of the pseudotensor current, to which we assign a dipole form factor. For the $p\to n$ transition, this is:

\begin{equation}
g^{pn}_2(Q^2) =  \frac{g_2(0)}{\left(1+\frac{Q^2}{M_A^2}\right)^2}    
\end{equation}

The relations between this form factor and its counterparts in hyperon production are the same as those for $g_1$ shown in table 1. We will explore the effects of a nonzero ``pseudotensor charge", $g_2(0)$, allowing it to have real and imaginary (time reversal violating) values.

SU(3) quark flavour symmetry violations are implemented through modification of the vector and axial vector form factors:

\begin{align}
f_1(Q^2) \to a f_1(Q^2) \\
g_1(Q^2) \to b g_1(Q^2)
\end{align}

The values of the coefficients $a$ and $b$ are calculated from a relativistic quark model by Schlumpf \cite{Schlumpf:1994fb} and are listed in table \ref{Table2}.

\begin{table}[H]
    \centering
    \begin{tabular}{|c|c|c|}
    \hline
        Process & $a$ & $b$  \\
        \hline
       $p \to \Lambda$  & 0.976 & 1.072 \\
       $p \to \Sigma^0$ & 0.975 & 1.051 \\
       $n \to \Sigma^-$ & 0.975 & 1.056 \\
       \hline
    \end{tabular}
    \caption{Symmetry breaking corrections from \cite{Schlumpf:1994fb}.}
    \label{Table2}
\end{table}

\subsection{Hyperon Potential}

The initial nucleon energy is modified by binding energy calculated from one of several models described in \cite{Juszczak:2005wk}. It is consistent with \cite{Sobczyk:2019uej} to include a potential for the outgoing hyperon of the form:

\begin{align}
V(r) = -\alpha \frac{\rho(r)}{\rho(0)}    
\end{align}

Where $\alpha=30$ MeV and $\rho(r)$ is the density of nuclear matter at a distance $r$ from the centre of the nucleus~\cite{DeJager:1987qc}. We use the following sign convention: $\alpha > 0$ is an attractive potential, $\alpha < 0$ is a repulsive potential.

\begin{comment}

Before propagation through the intranuclear cascade, the hyperon is returned to its mass shell: The 3 momentum is kept fixed, the effective mass $M_{Y}^*$ is replaced by the true mass and hyperon energy is recalculated. The hyperon then undergoes secondary interactions on shell.

This potential is incorporated into the cascade through hyperon absorption: At each step of propagation the kinetic energy of the hyperon $E_k$ in the nuclear rest frame is compared to $V(r)$ at its current position and if $V(r) > E_k$ the propagation of the hyperon is stopped and it does not escape the nucleus.

\end{comment}

In NuWro, this potential is incorporated into the intranuclear cascade in the following way: When the hyperon is initially inserted into the cascade, its total energy is increased by the value of the hyperon nucleus potential at that position, $V_0$ and its momentum adjusted accordingly to propagate it on shell. To accommodate repulsive potentials (in which some energy is \textit{subtracted} during this procedure), an additional check is made to ensure the hyperon can be moved into the potential while generating the primary interaction kinematics, the process is forbidden by the kinematics if not. At the end of every propagation step, the hyperon's kinetic energy $E_k$ is compared with $V_0$, if $E_k < V_0$, the hyperon is said to have been reabsorbed by the nucleus and propagation is ceased. At the end of the propagation process, $V_0$ is subtracted back off the total energy of the hyperon. 

If a secondary interaction occurs in which the hyperon switches families, ie from a $\Lambda$ to a $\Sigma$ or vice versa, the hyperon is moved to the new potential:  First the kinetic energy of the new hyperon is compared with the difference in the potentials and if $E_k < V_{\textrm{old}}-V_{\textrm{new}}$ this secondary interaction is ignored. Then the difference between the two potentials is subtracted from the new hyperon's energy and its momentum adjusted to continue propagating it on shell. This respects overall energy conservation.

\section{Intranuclear Cascade}
\label{Section3}

Once produced the hyperon has to be propagated through the nucleus and can rescatter off nucleons modifying the final state. The approach employed by NuWro to model this process is described in detail in Refs.~\cite{Golan:2012wx, Niewczas:2019fro}. A brief outline of the algorithm for propagating a particle is as follows:

\begin{enumerate}
    \item If the particle is not a nucleon, pion or hyperon, it exits the nucleus without interacting.
    \item Compute the mean free path $\lambda$ of the particle in the nucleus. Generate a step length $L$ for the particle from this mean free path using $L = -\lambda \ln{\xi}$ where $\xi \in [0,1]$ is a uniformly distributed random number. 
    \item The particle is propagated by a distance $d={\rm min}(L, 0.2~{\rm fm})$. This procedure accounts for the fact that $L$ depends on nuclear density and is calculated locally. If $L>0.2$~fm the particle is moved by a distance $0.2$~fm along its momentum vector and step 2. is repeated. 
    
    \item If $d<0.2$~fm and particle is not outside nucleus an interaction is generated.  A reaction is selected by MC method, comparing the cross sections of different reactions. The reaction is only simulated if final states of any nucleons are not Pauli blocked.
\end{enumerate}

This cascade requires in medium hyperon-nucleon cross sections. In this paper the fits to free hyperon-nucleon data from~\cite{Singh} are used for total cross sections. 

The differential cross sections have been assigned the same distributions as $NN$ scattering processes with corresponding electric charges. For example: 

\begin{align}
\frac{d\sigma}{d\Omega}(p \Lambda \to p \Lambda)\sim
    \frac{d\sigma}{d\Omega}(pn \to pn) 
\end{align}

In cases where no similar process in the $NN$ sector exists (e.g. for $\Sigma^-$ rescatterings), the outgoing particles are scattered isotropically in the CMS frame. 
The arguments justifying this assumption are that the underlying physics responsible for the interactions, the strong force, is the same and the energy is small enough that any differences in the internal structure of the particles, i.e. the parton distributions, are negligible.  The differential cross sections are then obtained from \cite{Metropolis:1958wvo} and have the form:

%\pur{a parametrization of Ref.~\cite{Cugnon:1996kh}, in} \purs{Metropolis et al\cite{Metropolis:1958wvo} and have} the form

%\begin{align}
%\frac{d\sigma}{d\Omega} = K \left( A\cos^4{\theta} + B\cos^3{\theta} + C\cos{\theta} + 1 \right),   
%\end{align}

\begin{align}
\frac{d\sigma}{d\Omega} = K \left( A\cos^4{\theta} + B\cos^3{\theta} + 1 \right),   
\end{align}

where $K$ is a normalisation constant, while $A$ and $B$ are fitted parameters dependant on the hyperon kinetic energy and $\theta$ is the scattering angle in the CMS frame. 

\section{Free Nucleon Case}
\label{Section4}

\subsection{Total Cross Sections}

We study the sensitivity of the hyperon cross section to the previously described second class current and SU(3) symmetry violations, and variations in the axial mass. In figures 1-4 we compare the total cross section prediction for $\Lambda$ production from free nucleons using muon antineutrinos of energies up to 15 GeV, covering the full range of energies used by the experimental fluxes studied in section VI excluding the one used by the Neutrino Oscillation Magnetic Detector (NOMAD) experiment~\cite{Astier:2003rj}. 

\begin{figure}[H]
    \centering
    \includegraphics[width=\linewidth,height=8.7cm]{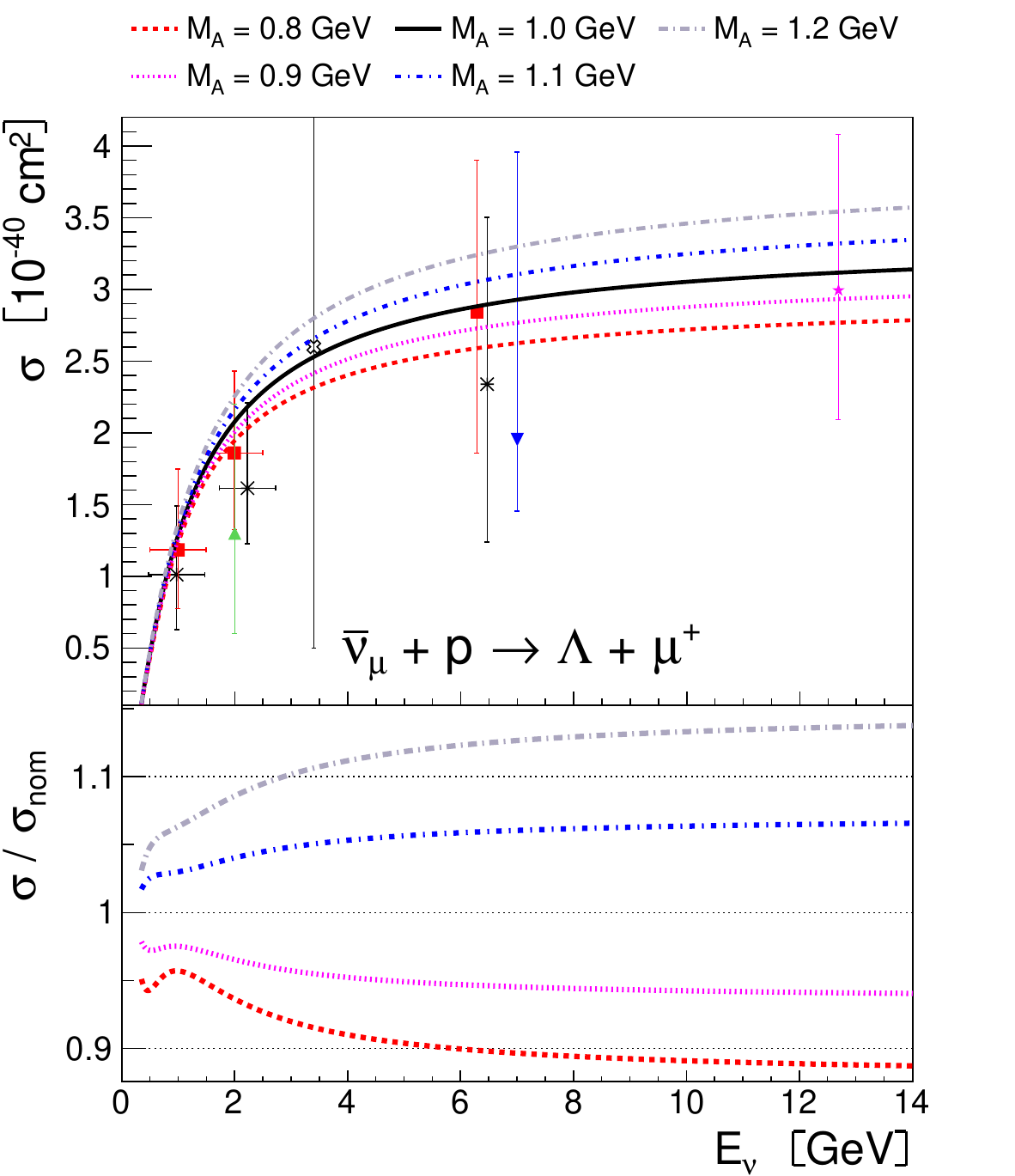}
    \caption{Total cross section for $\Lambda$ production on free protons for several axial masses. Bottom shows the ratio of the cross section for some values of $M_A$ to the cross section at $M_A=1$ GeV. Data is taken from Ammosov\cite{Ammosov:1986jn} (FNAL Bubble Chamber, pink star), Brunner\cite{Brunner:1989kw} (SKAT, triangle down), Erriquez\cite{Erriquez1,Erriquez2} (Gargamelle, red square and black x respectively), Eichten\cite{Eichten:1972bb} (Gargamelle, triange up) and Fanourakis\cite{Fanourakis:1980} (BNL Bubble Chamber, white cross).}
    \label{AntiMu_Proton_Lambda_XSec_MA}
\end{figure}

\begin{figure}[H]
    \centering
    \includegraphics[width=\linewidth,height=8.7cm]{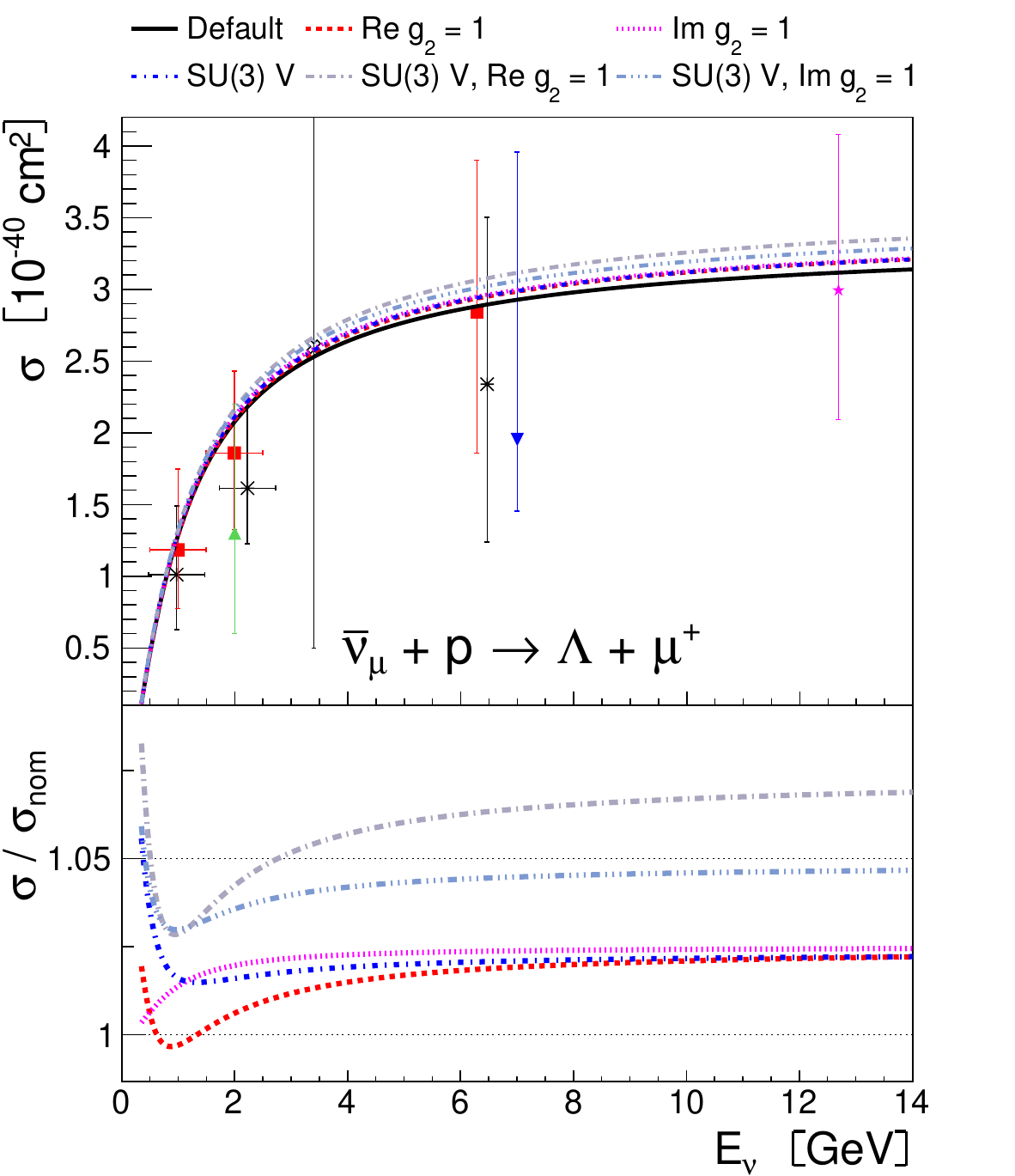}
    \caption{ Total cross section for $\Lambda$ production from free protons for several configurations compared to experimental data described in figure \ref{AntiMu_Proton_Lambda_XSec_MA}. Bottom is the ratio of cross sections to the nominal (no SCC or SU(3)V).}
    \label{AntiMu_Proton_Lambda_XSec_BSM}
\end{figure}

A large range of hyperon axial masses are shown, consistent with the spread of measurements of the CCQE axial mass from past experiments \cite{Erriquez1,Miniboone1}. The resulting predictions are comfortably within the error bars of the existing hyperon production data. The relative change in the cross section with respect to the default model ($M_A=1.0$ GeV, no SCC or SU(3)V) is shown in the bottom panels of figures 1-4 for these axial masses and SCC/SU(3)V configurations. Comparing these curves in figures \ref{AntiMu_Proton_Lambda_XSec_MA} and \ref{AntiMu_Proton_Lambda_XSec_BSM} indicates much larger cross section variations are produced by the changes to axial mass; if this parameter is poorly constrained the second class current and symmetry breaking effects may be difficult to measure. These effects are greatest for large neutrino energies, above 5 GeV, larger those produced by the Booster neutrino~\cite{AguilarArevalo:2008yp} and NOvA beams~\cite{Acero:2019ksn} but within the reach of future experiments such as DUNE \cite{Abi:2020qib}.

\begin{figure}[H]
    \centering
    \includegraphics[width=\linewidth,height=8.7cm]{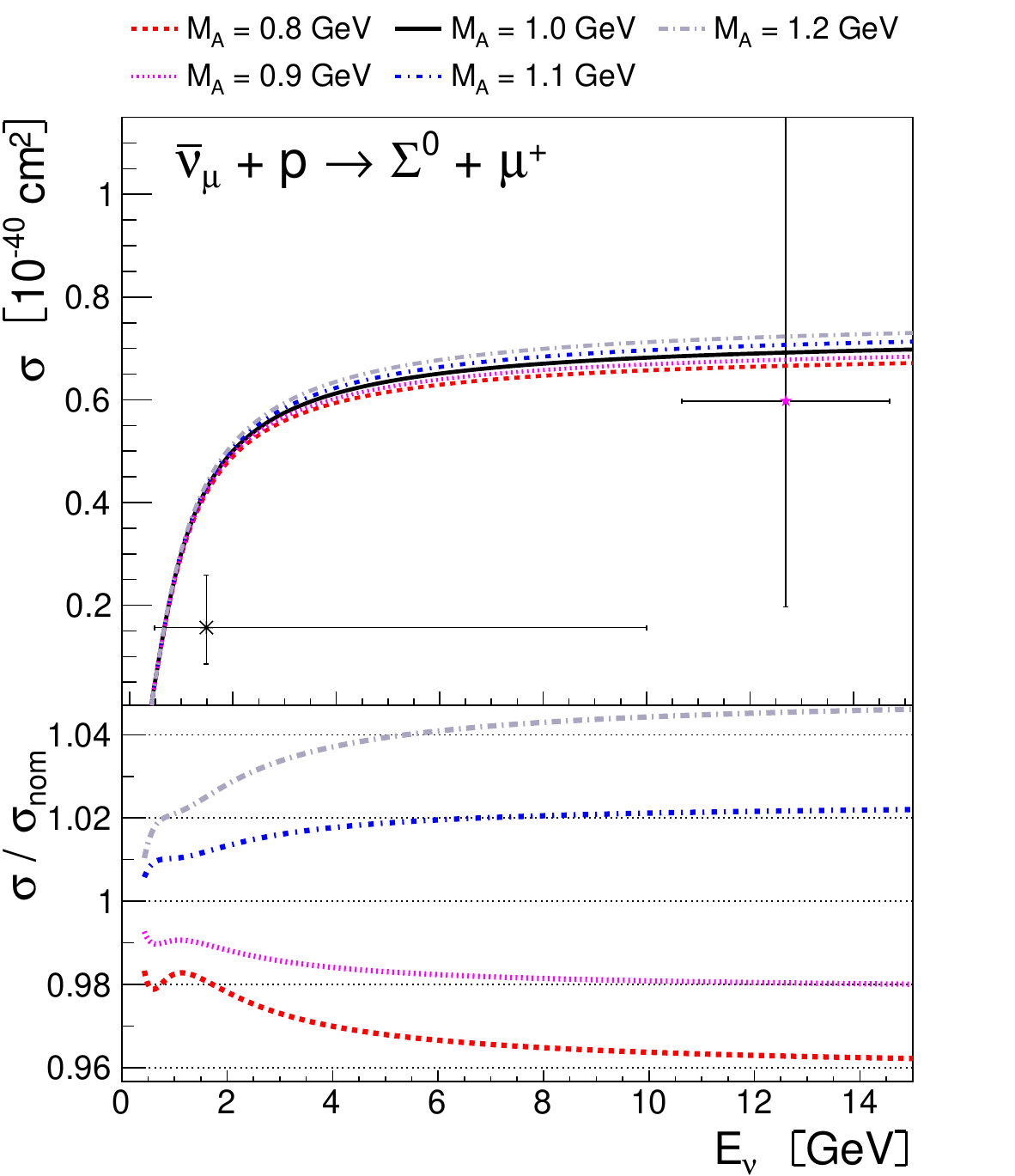}
    \caption{Total cross section for $\Sigma^0$ production from free protons for several axial masses. Bottom is the ratio of cross section for some values of $M_A$ to the cross section at $M_A=1$ GeV.}
    \label{AntiMu_Proton_Sigma_XSec_MA}
\end{figure}

\begin{figure}[H]
    \centering
    \includegraphics[width=\linewidth,height=8.7cm]{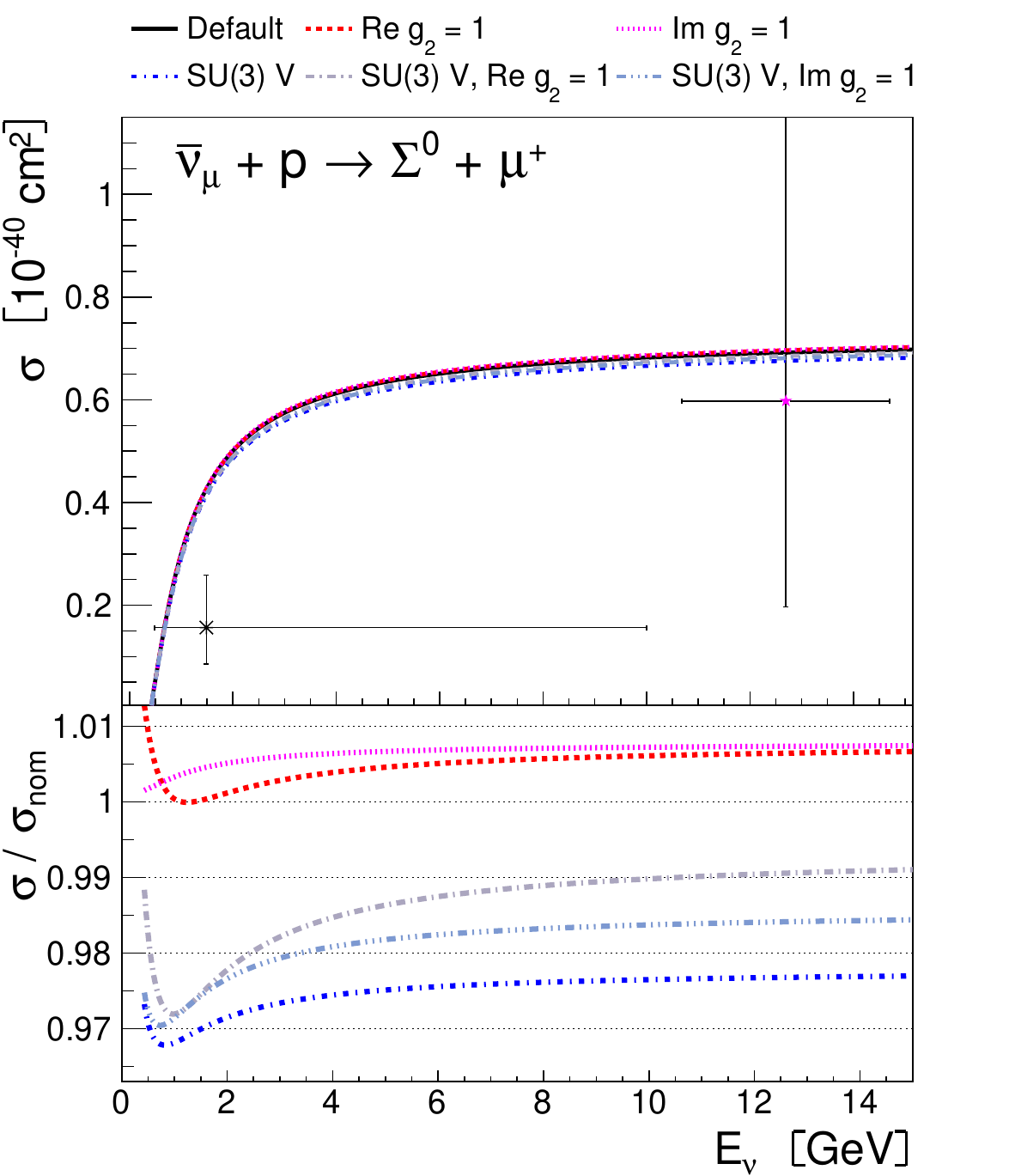}
    \caption{Total cross section for $\Sigma^0$ production from free protons for several configurations compared to experimental data as described above. Bottom is the ratio of cross sections to the nominal (no SCC or SU(3)V).}
    \label{AntiMu_Proton_Sigma_XSec_BSM}
\end{figure}

The predictions for $\Sigma^0$ production are compared to data in figures \ref{AntiMu_Proton_Sigma_XSec_MA} and \ref{AntiMu_Proton_Sigma_XSec_BSM}. The axial mass affects this cross section in a similar way to that of the $\Lambda$ however the overall effect is weaker, a consequence of the smaller SU(3) coefficient of $g_A$ shown in table I. The cross section is decreased by the symmetry breaking corrections, the opposite effect to that seen for the $\Lambda$. To explore this further we consider the ratio:

\begin{align}
R = \frac{\sigma(\bar{\nu}_l + p \to \Lambda + l^+)}{\sigma(\bar{\nu}_l + p \to \Sigma^0 + l^+)  + \sigma(
\bar{\nu}_l +n \to \Sigma^- + l^+)}    
\end{align}

\begin{figure}[H]

    \centering
    \includegraphics[width=\linewidth,height=5.4cm]{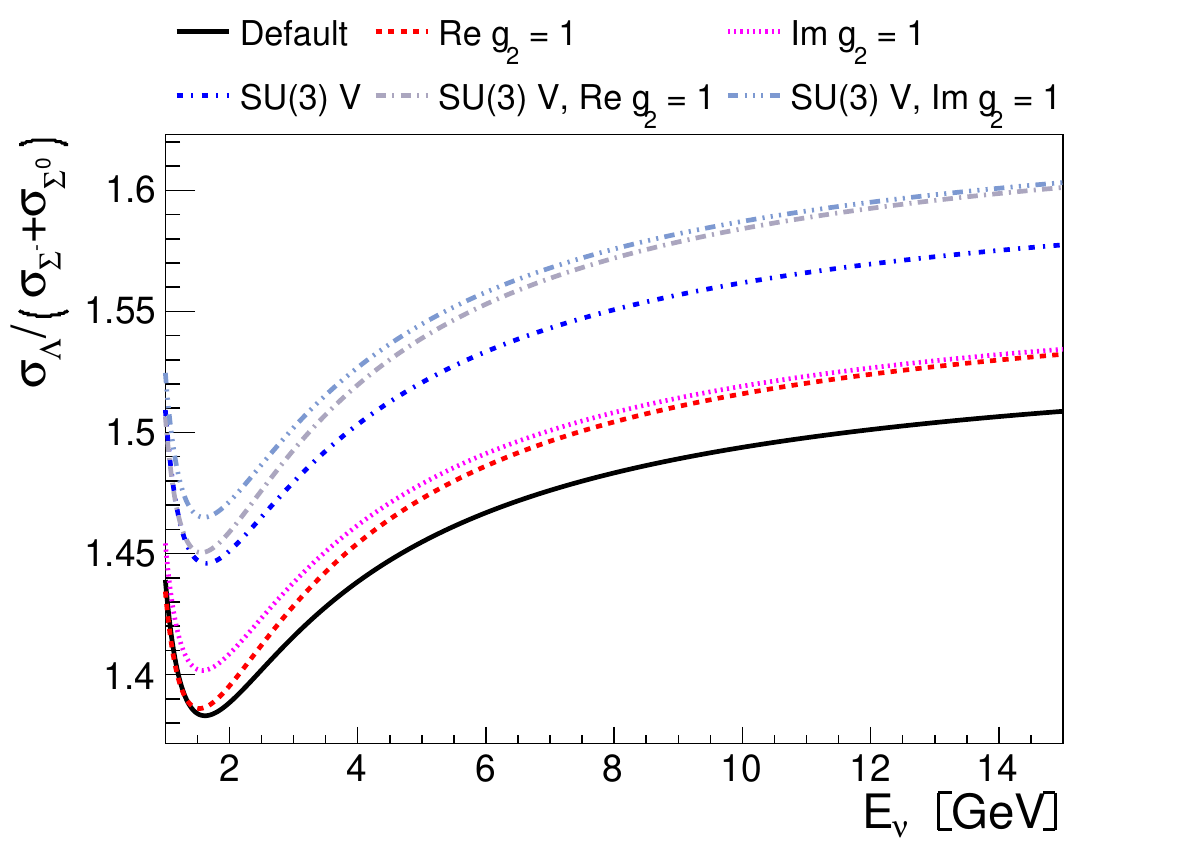}
    \caption{Ratio of $\Lambda$ to $\Sigma^{0,-}$ cross sections described by equation 25.}
    \label{Lambda_Sigma_Ratio_BSM}
\end{figure}

The values of this ratio for different SCC/symmetry breaking configurations are shown in figure \ref{Lambda_Sigma_Ratio_BSM} with the relative change shown in the bottom panel. This ratio increases for all the settings shown for almost the entire energy range and the relative change is roughly constant for neutrino energies above a few GeV, however this change is not significantly larger than that in the $\Lambda$ cross section.

\newpage

%\begin{figure}[H]
%    \centering
%    \includegraphics[width=0.91\linewidth]{Lambda_Sigma_Ratio_MA.pdf}
%    \caption{Ratio of $\Lambda$ to $\Sigma^{0,-}$ described by equation 25. The bottom panel is the ratio of $R$ using several axial mass values to its value at $M_A = 1$ GeV.}
%\end{figure}

\section{Nuclear Effects}
\label{Section5}
\subsection{Differential Cross Sections}

To study the effects of FSI we compare the differential cross sections for production of the different hyperon species before and after application of the cascade for three beam energies, shown in figures 6-10. The differential cross section after FSI, shown as a dashed line, is calculated from the fraction of hyperon events in which a hyperon exists in the final state. We study the effect of FSI on several variables, including $Q^2$ and outgoing hyperon kinetic energy. Since interactions between the outgoing lepton and the nucleus are not simulated, the $Q^2$ of an individual interaction is unmodified, however if one studies exclusive channel cross sections, the $Q^2$ distribution changes as a result of hyperon conversions and absorption. 

\begin{figure}[H]
    \centering
   \includegraphics[width=\linewidth,height=5.4cm]{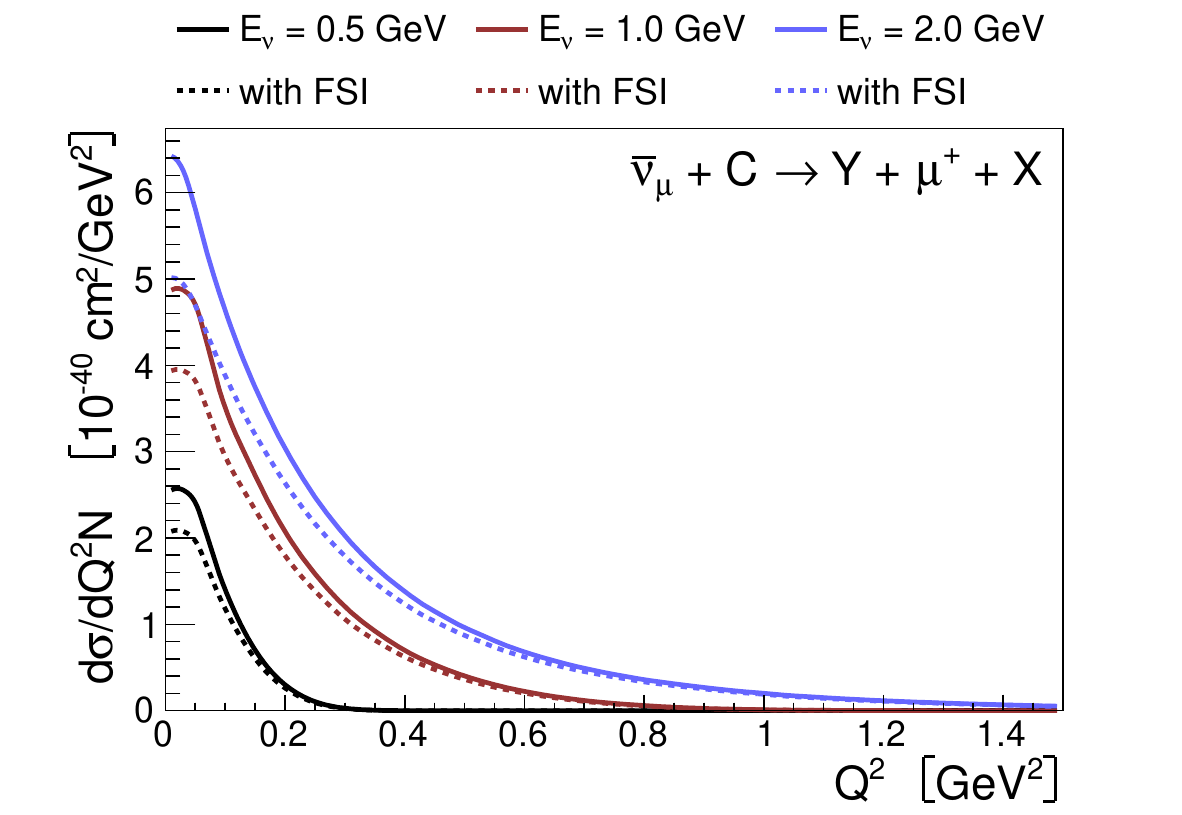}
    \caption{Inclusive hyperon production differential cross section per nucleon for $\bar{\nu}_{\mu} \textrm{C} \to Y \mu^+ +X$ at three neutrino energies. $X$ denotes  any additional final state particles. Solid lines show the differential cross section before propagation of the hyperon through the cascade, dashed lines are after.}
    \label{Basic_C_Incl_Q2}
\end{figure}

The presence of the hyperon potential is seen in figure \ref{Basic_C_Incl_Q2} through the quenching of the cross section, which is strongest in low $Q^2$ events. In this section a local Fermi gas model is used to describe the initial nucleon state. Comparison of figures \ref{Basic_C_Sigma_Q2} and \ref{Basic_C_SigmaM_Q2} shows relation (16) is approximately maintained after final state interactions.

Carbon and argon are common target materials in neutrino experiments such as carbon in NOvA and MiniBooNE  and argon in SBND and DUNE; the inclusive hyperon production cross sections for targets are compared in figure \ref{Basic_C_vs_Ar_Incl_Q2}. We expect a larger cross section per nucleon from carbon: This can be understood as the combined cross sections for $\Lambda$ and $\Sigma^0$ from protons are larger than the cross section for $\Sigma^-$ from neutrons and thus a target with a greater fraction of mass from protons will yield a larger cross section per nucleon. The effect of the larger argon nucleus can be seen through the difference between the post-FSI differential cross sections, which differ by a far greater amount than the pre-FSI curves.

\begin{figure}[H]
    \centering
    \includegraphics[width=\linewidth,height=5.4cm]{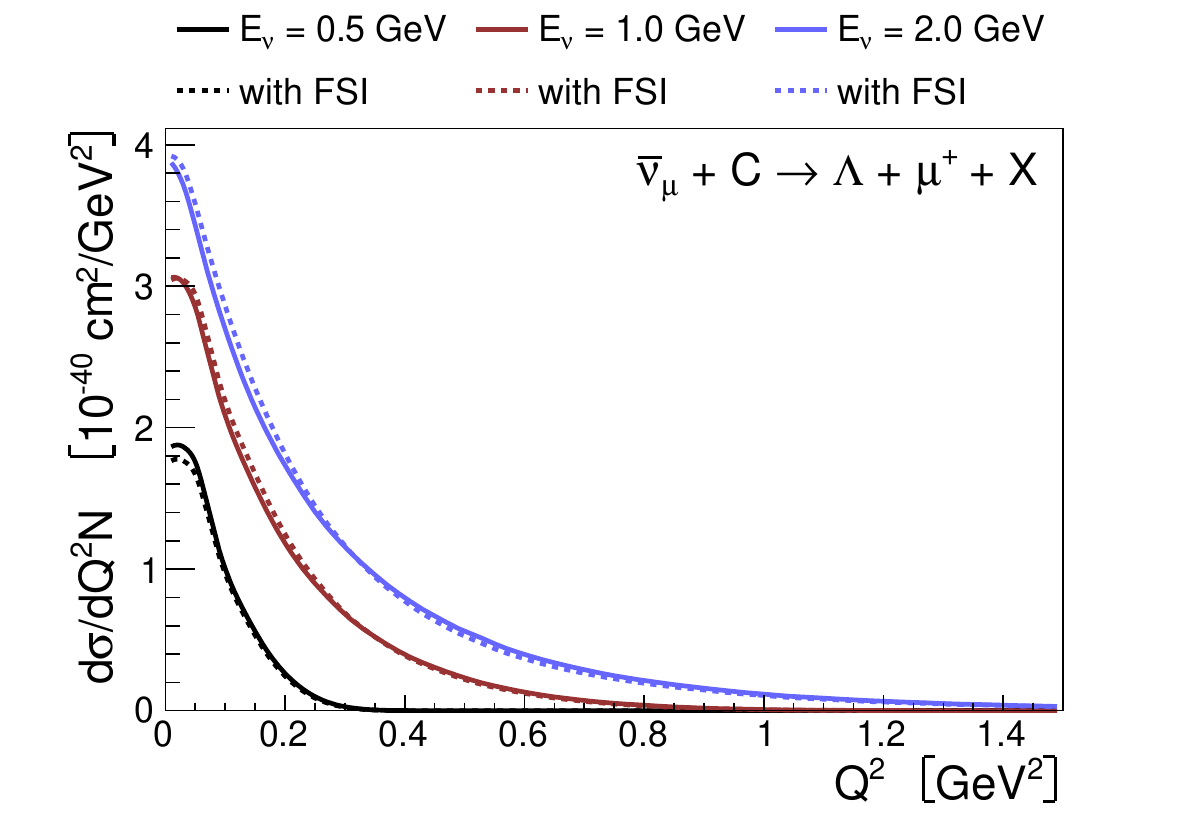}
    \caption{Differential cross section per nucleon for $\bar{\nu}_{\mu} \textrm{C} \to \Lambda \mu^+$ at three neutrino energies. }
    \label{Basic_C_Lambda_Q2}
\end{figure}

\begin{figure}[H]
    \centering
    \includegraphics[width=\linewidth,height=5.4cm]{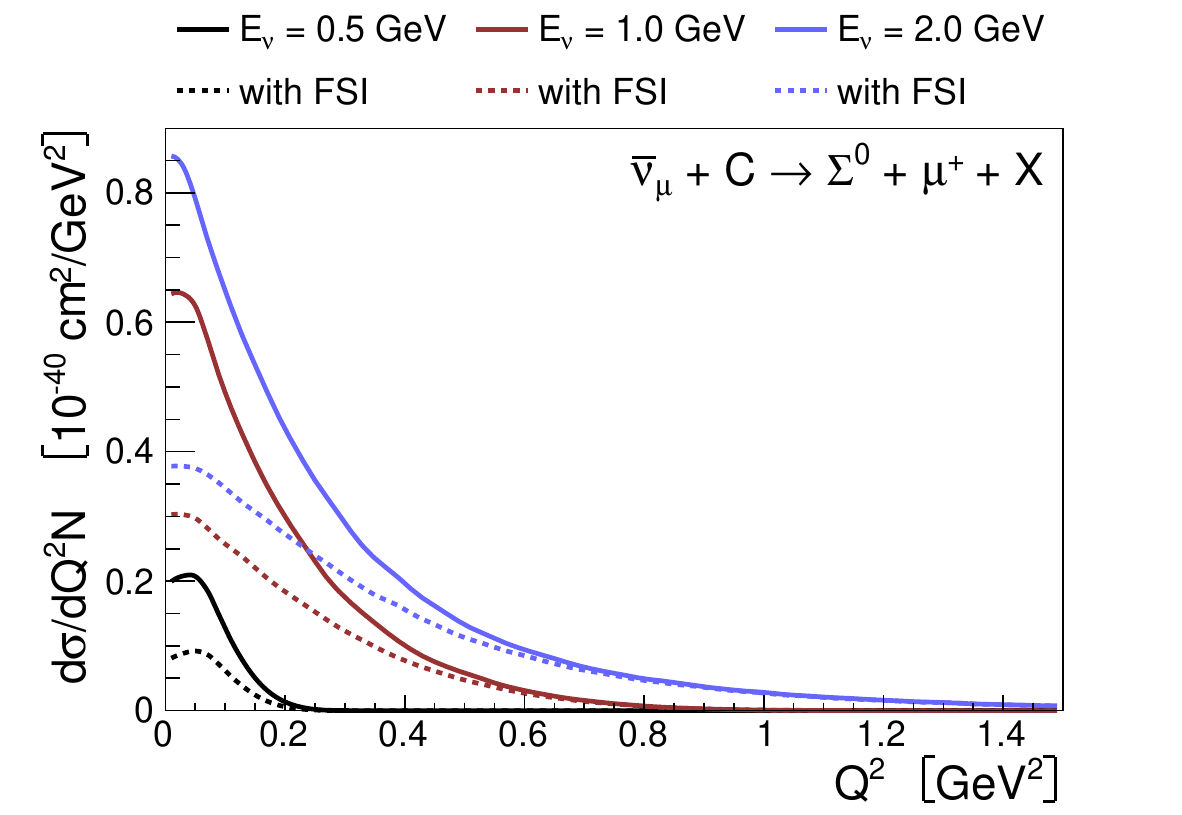}
    \caption{Differential cross section per nucleon for $\bar{\nu}_{\mu} \textrm{C} \to \Sigma^0 \mu^+$ at three neutrino energies.}
    \label{Basic_C_Sigma_Q2}
\end{figure}

\begin{figure}[H]
    \centering
    \includegraphics[width=\linewidth,height=5.4cm]{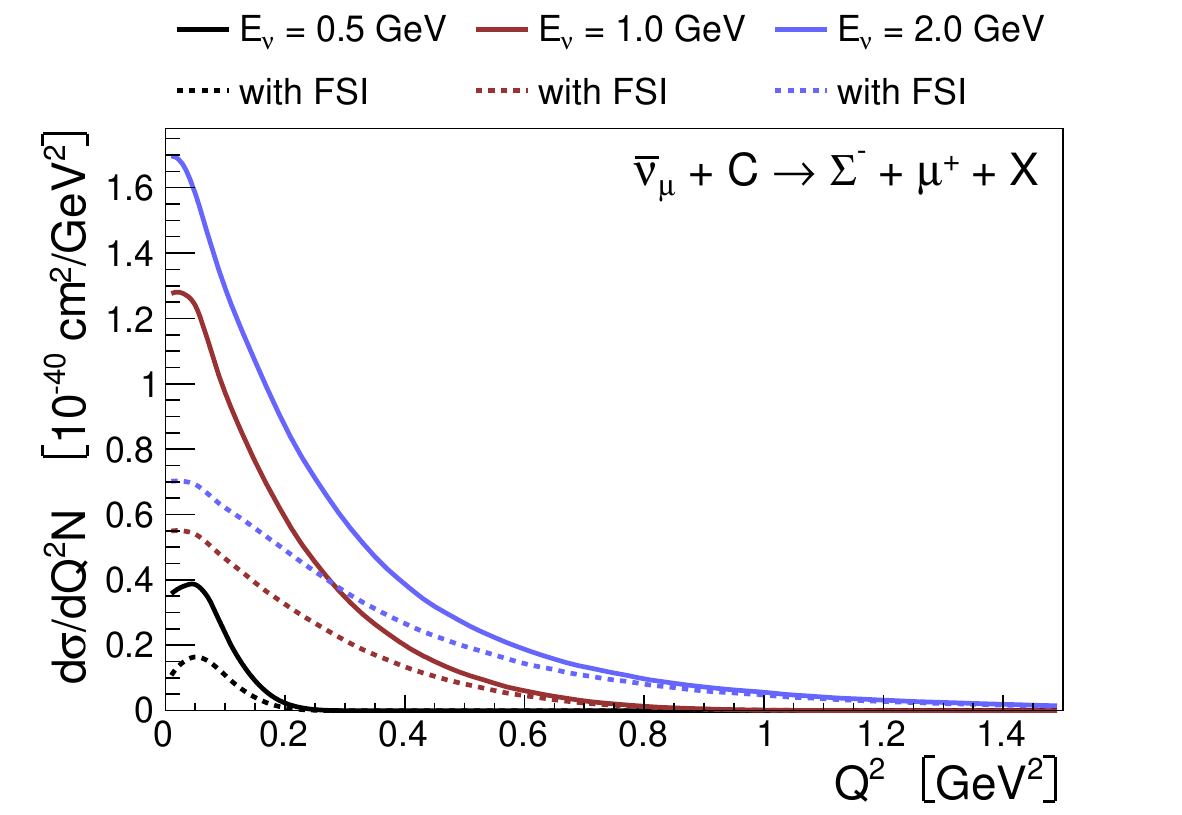}
    \caption{Differential cross section per nucleon for $\bar{\nu}_{\mu} \textrm{C} \to \Sigma^- \mu^+$ at three neutrino energies.}
    \label{Basic_C_SigmaM_Q2}
\end{figure}

\begin{figure}[H]
    \centering
    \includegraphics[width=\linewidth,height=5.4cm]{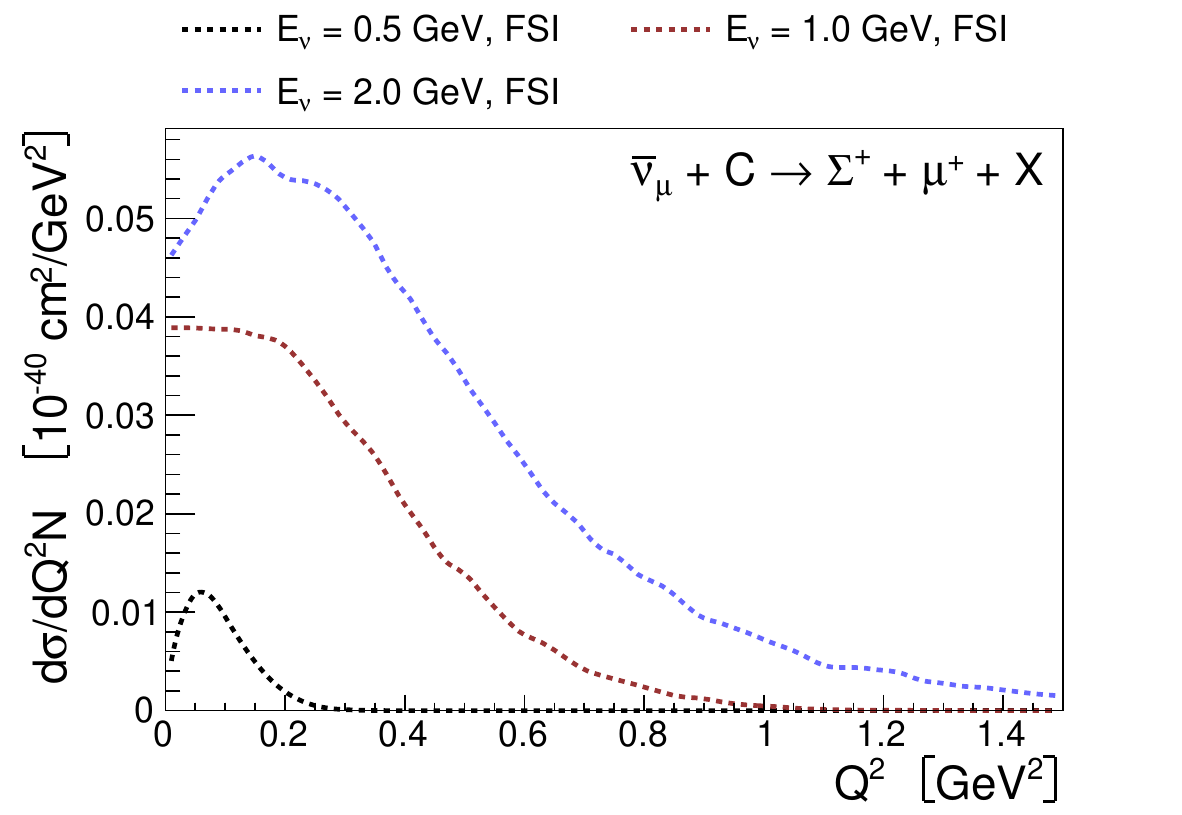}
    \caption{Differential cross section per nucleon for $\bar{\nu}_{\mu} \textrm{C} \to \Sigma^+ \mu^+$. The $\Sigma^+$ can only be produced through FSI thus no pre-FSI (solid) curves are shown.}
    \label{Basic_C_SigmaP_Q2_FSI}
\end{figure}

\begin{figure}[H]
    \centering
    \includegraphics[width=\linewidth,height=5.4cm]{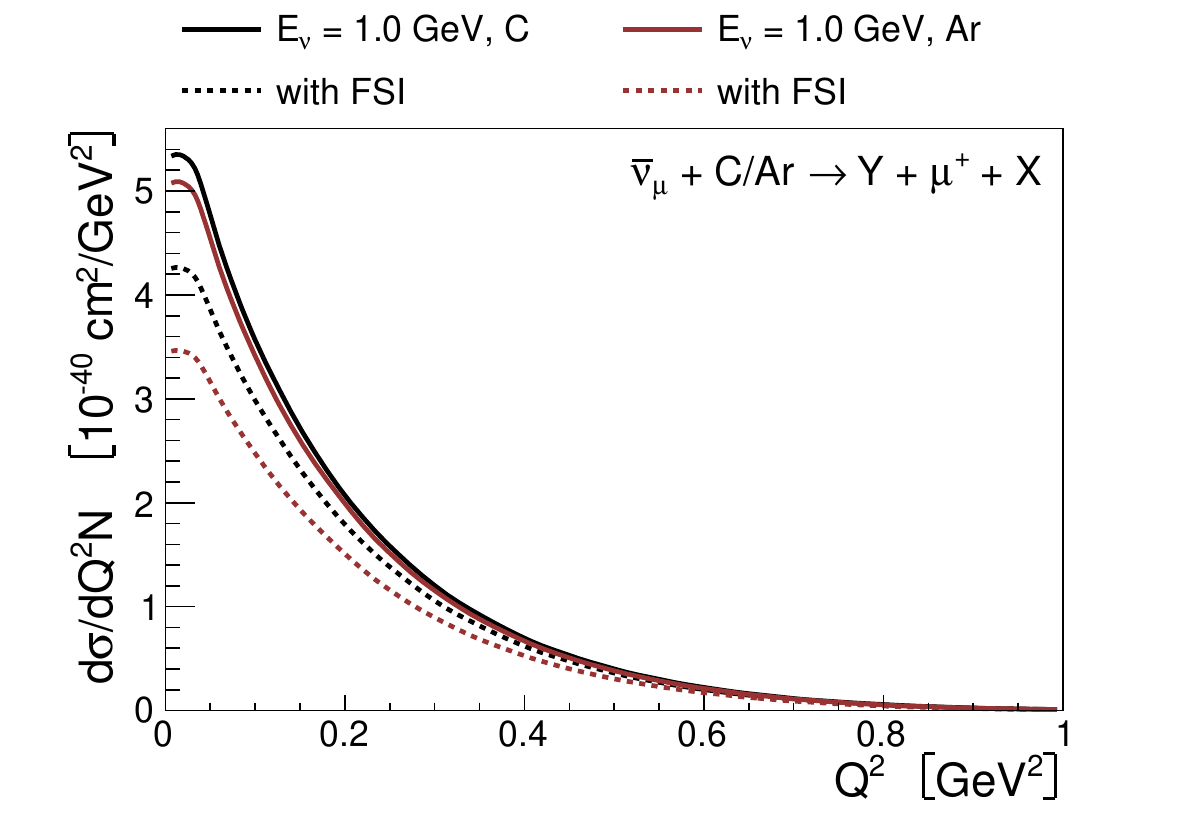}
    \caption{Inclusive differential cross section per nucleon for using C and Ar as targets.}
    \label{Basic_C_vs_Ar_Incl_Q2}
\end{figure}

\subsection{$\Lambda$ Enhancement}

Rescattering in this work conserves the total number of hyperons, but allows them to transform from one type to another. The cross section fits for interactions between nucleons and hyperons often contain kinematic factors that suppress the conversion of $\Lambda$ into $\Sigma$, whilst the inverse processes receive an enhancement. This is the most significant of the nuclear effects. 

\begin{figure}[]
    \centering
    \includegraphics[width=\linewidth,height=5.4cm]{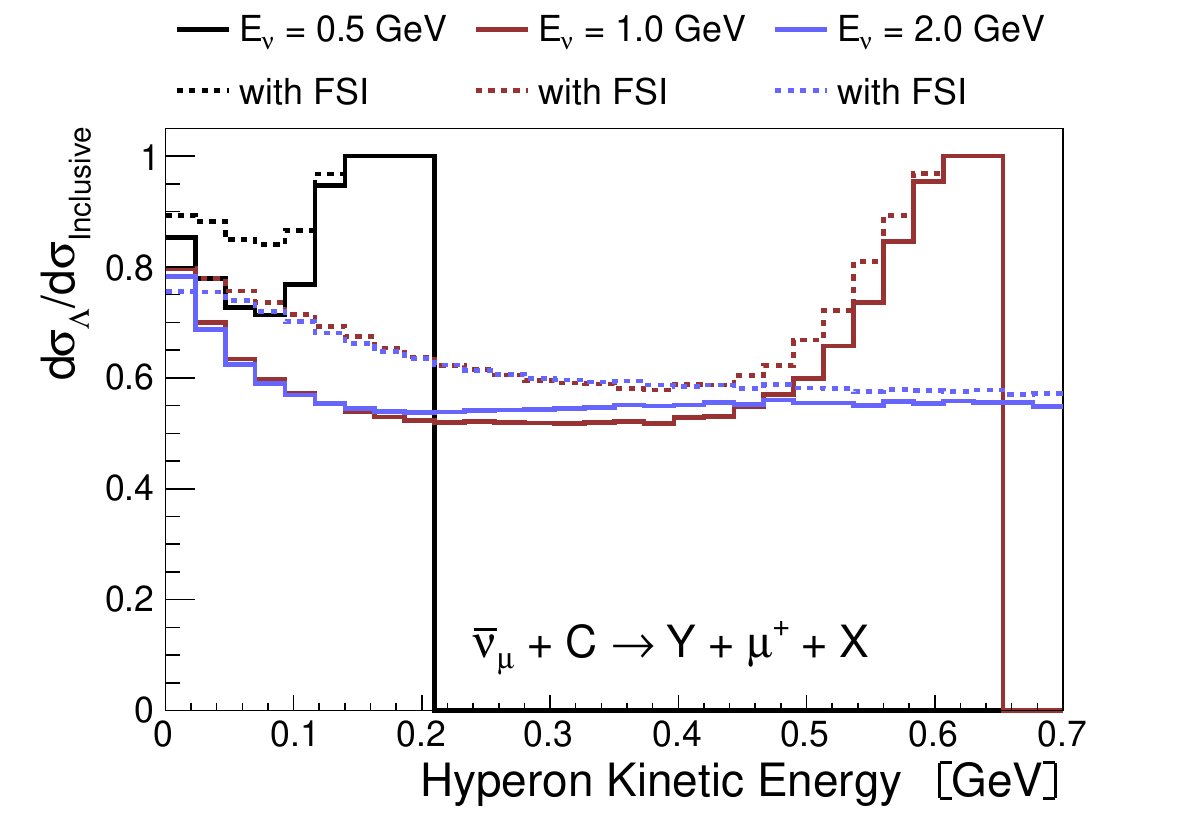}
    \caption{Ratio of differential cross section for $\Lambda$ production to the inclusive differential cross section for several neutrino energies.}
    \label{Basic_C_Lambda_Hyperon_KE_Ratio}
\end{figure}

Comparison of figure \ref{Basic_C_Lambda_Q2} with figures \ref{Basic_C_Sigma_Q2} and \ref{Basic_C_SigmaM_Q2} shows the application of FSI slightly increases the $\Lambda$ cross section, while the opposite effect is seen in the $\Sigma$ channels. To study this further, we calculate the ratio of the $\Lambda$ differential cross section to the inclusive differential cross section, before and after traversal of the cascade, shown in figure \ref{Basic_C_Lambda_Hyperon_KE_Ratio}. At all of the neutrino energies shown, the fraction of the cross section due to $\Lambda$ increases, approaching 1 as the hyperon kinetic energy increases to its maximum value; the $\Lambda$ can be produced with a slightly larger kinetic energy than the $\Sigma$'s.

\subsection{$\Sigma^+$ Production}

\begin{figure}[H]
    \centering
    \includegraphics[width=\linewidth,height=5.4cm]{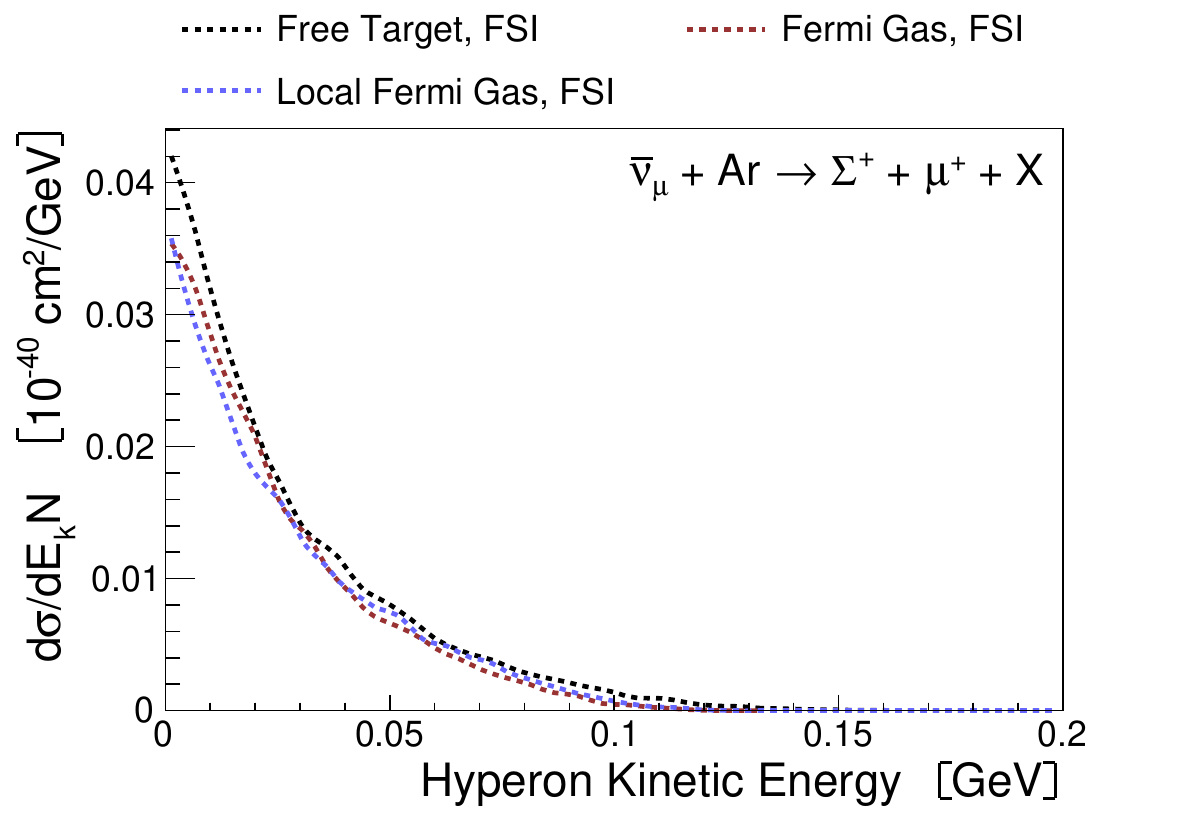}
    \caption{Differential cross section with respect to the hyperon kinetic energy for production of $\Sigma^+$ from argon at $E_{\nu} = 0.5$ GeV }
    \label{Nuclear_Models_Ar_SigmaP_Hyperon_KE_FSI}
\end{figure}

\begin{figure}[H]
    \centering
    \includegraphics[width=\linewidth,height=5.4cm]{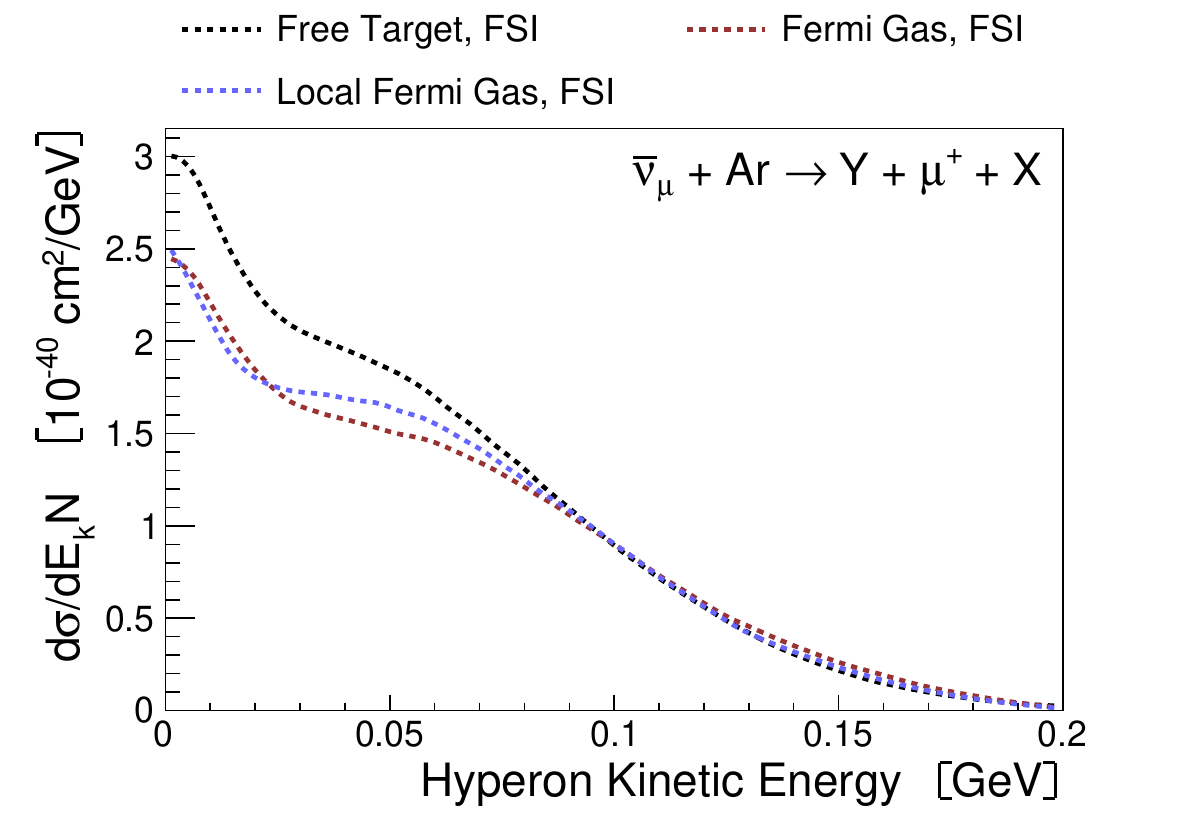}
    \caption{Inclusive differential cross section for hyperon production from argon at $E_{\nu} = 0.5$ GeV.}
    \label{Nuclear_Models_Ar_Incl_Hyperon_KE_FSI}
\end{figure}

$\Sigma^+$ baryons are produced exclusively through secondary interactions and their production rate and kinematics are expected to be very sensitive to nuclear effects. Figures \ref{Nuclear_Models_Ar_SigmaP_Hyperon_KE_FSI} and \ref{Nuclear_Models_Ar_Incl_Hyperon_KE_FSI} show the kinetic energy spectra of hyperons produced at $E_{\nu}=0.5$ GeV from Ar using three different nuclear models: A free target in which the nucleons are stationary, a global Fermi gas (GFG) where the nucleus is modelled as ball of constant density and a local Fermi gas (LFG) where a realistic density profile is used~\cite{DeJager:1987qc}.

The models produce similarly shaped $\Sigma^{+}$ kinetic energy distributions, with a small difference in the total cross section between the free target and GFG models. A significant change in the prediction occurs when switching to the LFG. The inclusive cross sections in figure \ref{Nuclear_Models_Ar_Incl_Hyperon_KE_FSI} separate at lower energies, when the reinteracting contribution becomes significant.

\begin{comment}
The difference between models is seen at low hyperon kinetic energies: Additional hyperons are produced at low energies when the free target model is used and the difference between the GFG and LFG is produces a shift in the peaks of these distributions. Comparing the inclusive distribution to the $\Sigma^+$ kinetic energies, the $\Sigma^+$s are typically produced with lower energies, this is expected given their production mechanism. 
\end{comment}

\subsection{Hyperon Nucleus Potentials}

There have been attempts to determine the hyperon nucleus potential using pion-nucleus scattering events in which a kaon is observed in the final state~\cite{Rodriguez-Sanchez:2018oqv,Harada:2005hs,Saha:2004ha}. These work suggest separate strengths for the $\Lambda$ and $\Sigma$ potentials. Here we examine the effects of varying the strengths of these potentials. The authors of~\cite{Rodriguez-Sanchez:2018oqv} suggest the following range for the $\Lambda$ nucleus potential for symmetric nuclei:

\begin{align}
\alpha_{\Lambda} \in \left[ 25 \textrm{ MeV} , 29\textrm{ MeV} \right]
\end{align}

The $\Sigma$ nucleus potential is less well constrained. \cite{Saha:2004ha} compares the range of strengths, in combination with $\alpha_{\Lambda} = 30$ MeV, allowing this potential to be strongly repulsive:

\begin{align}
\alpha_{\Sigma} \in \left[ 10 \textrm{ MeV} , -150 \textrm{ MeV} \right]
\end{align}

\begin{figure*}
\subfigure{\includegraphics[width=0.5\linewidth,height=4.9cm]{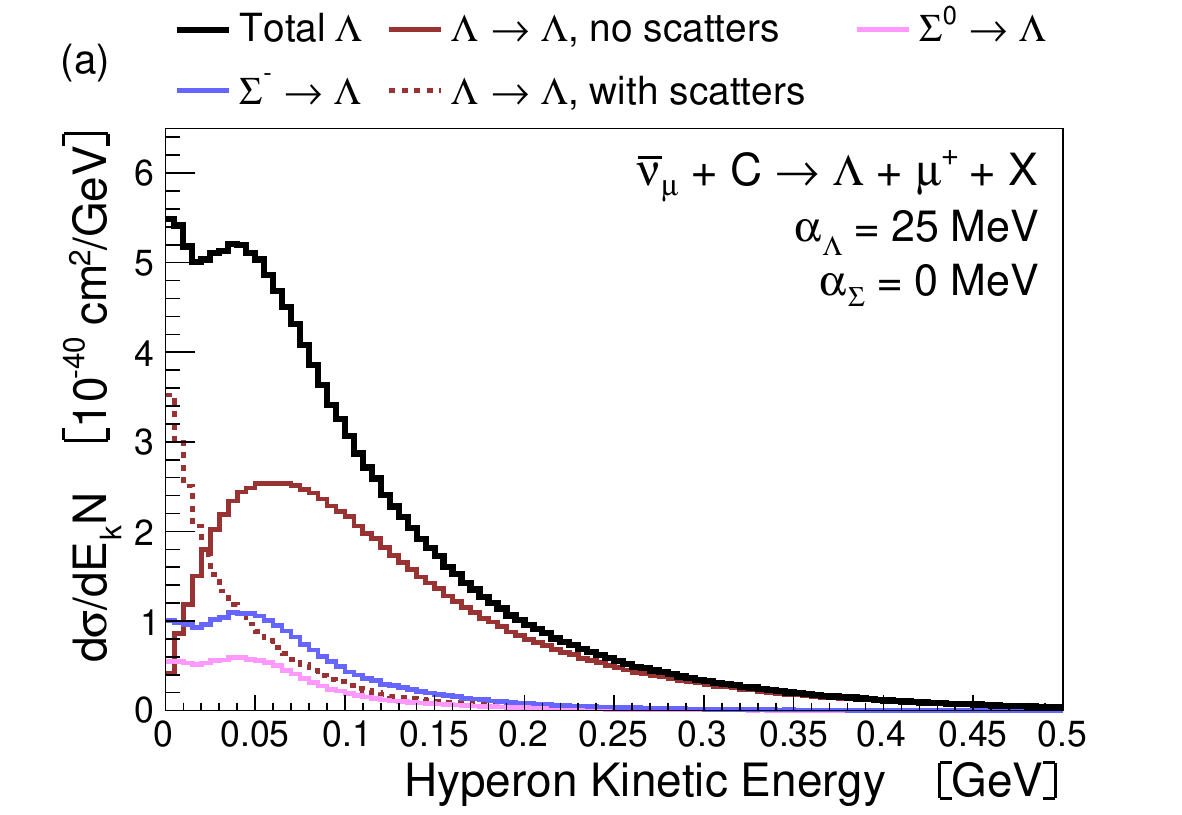}}%
\subfigure{\includegraphics[width=0.5\linewidth,height=4.9cm]{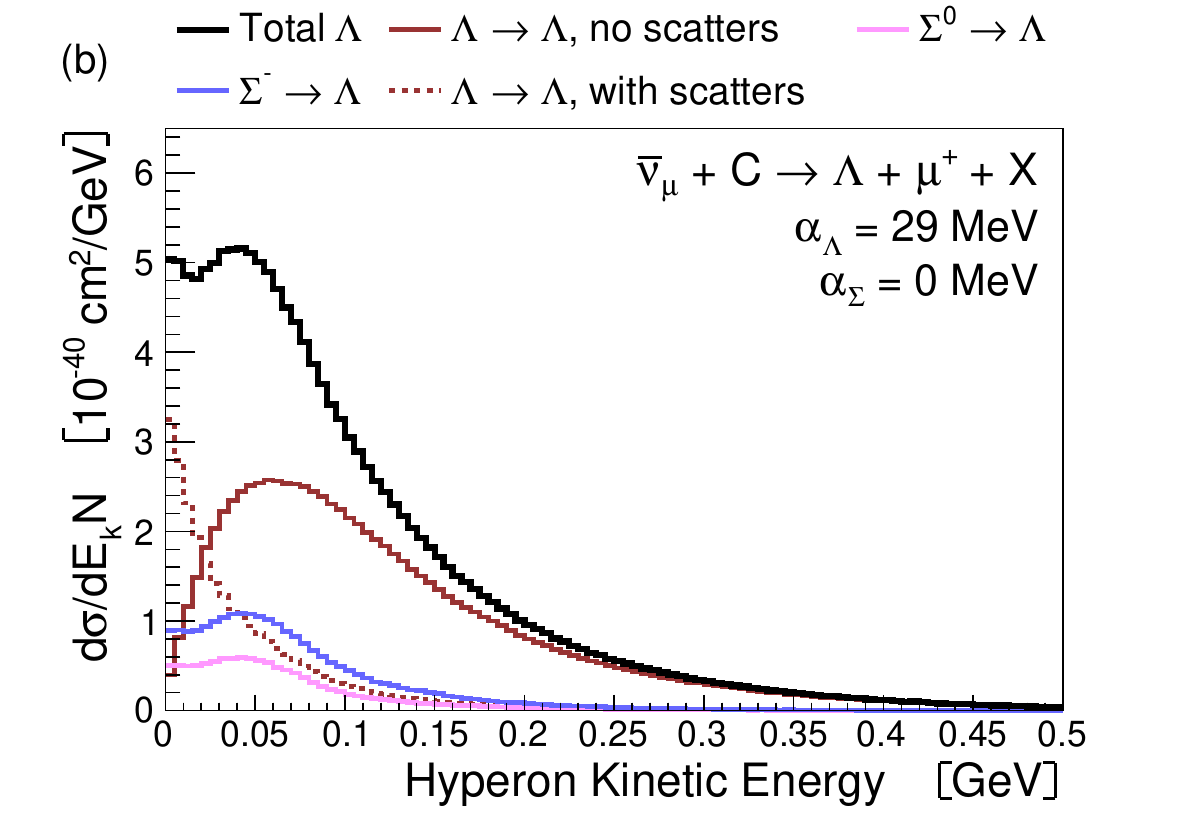}}
\subfigure{\includegraphics[width=0.5\linewidth,height=4.9cm]{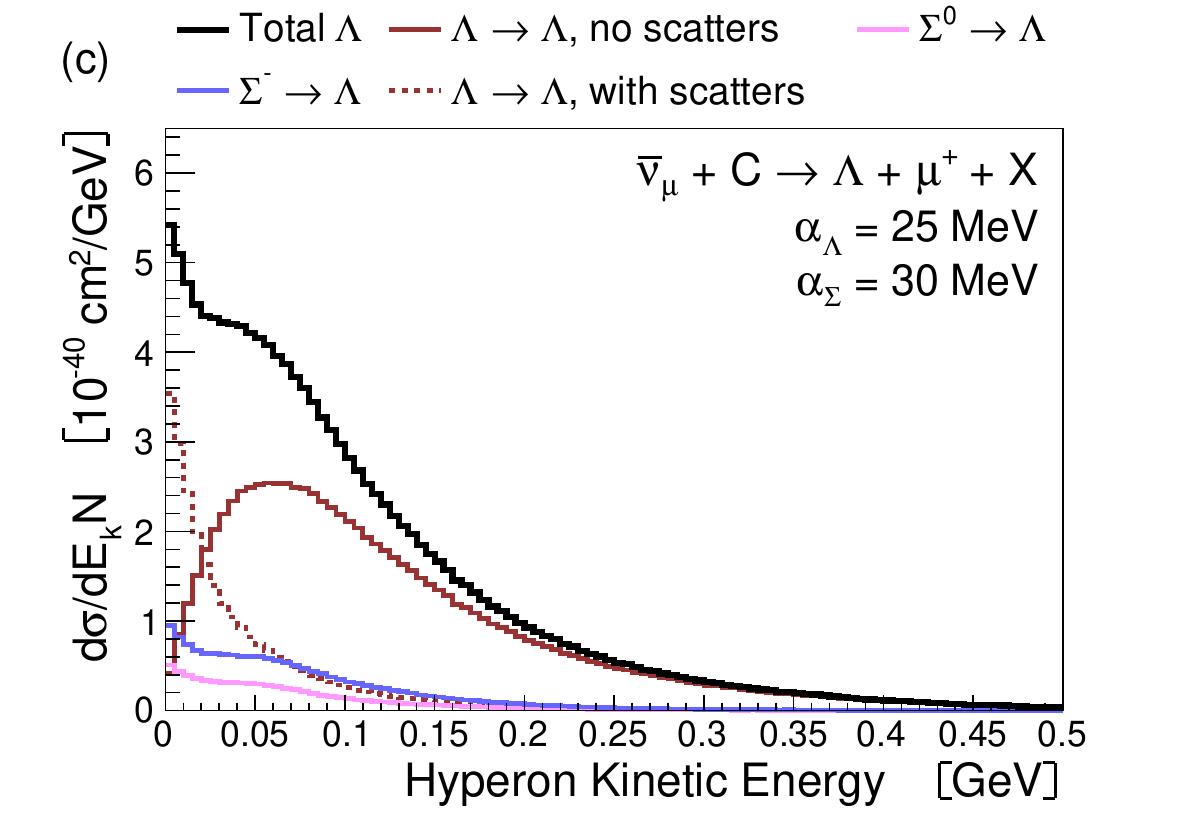}}%
\subfigure{\includegraphics[width=0.5\linewidth,height=4.9cm]{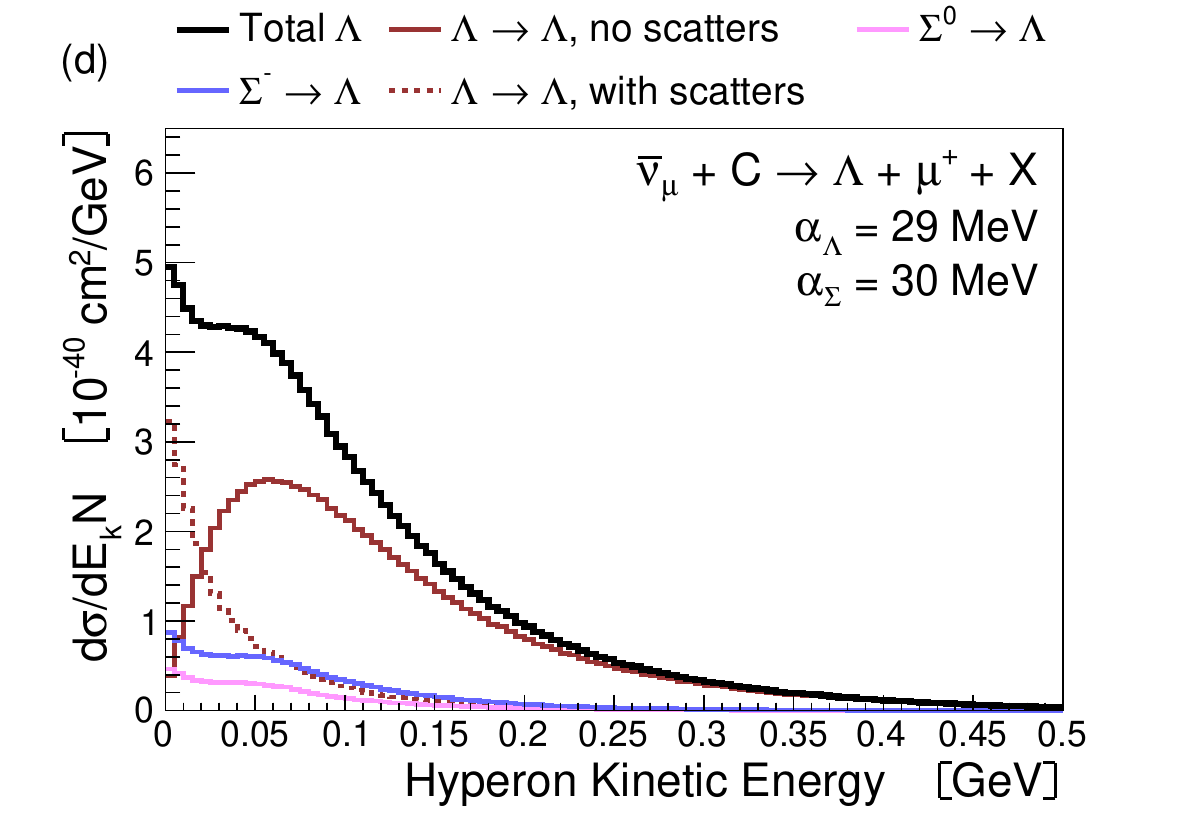}}
\caption{Distribution of $\Lambda$ kinetic energies after FSI for several choices of the hyperon-nucleus potentials. The total differential cross section is given by the solid black line, which can be broken down into several contributions, shown by the coloured lines.}
\label{UnStacked_Lambda_FSI_Rodriguez}
\end{figure*}

\begin{figure*}
\subfigure{\includegraphics[width=0.5\linewidth,height=4.9cm]{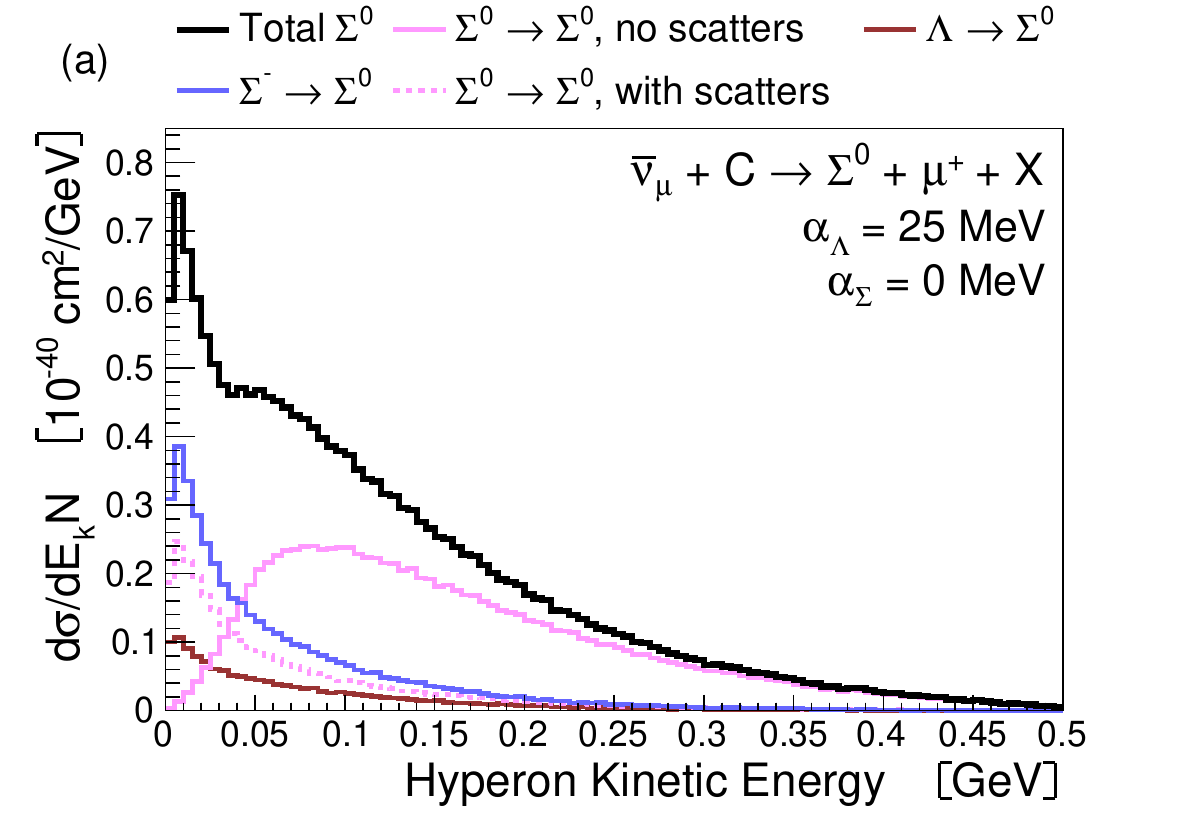}}%
\subfigure{\includegraphics[width=0.5\linewidth,height=4.9cm]{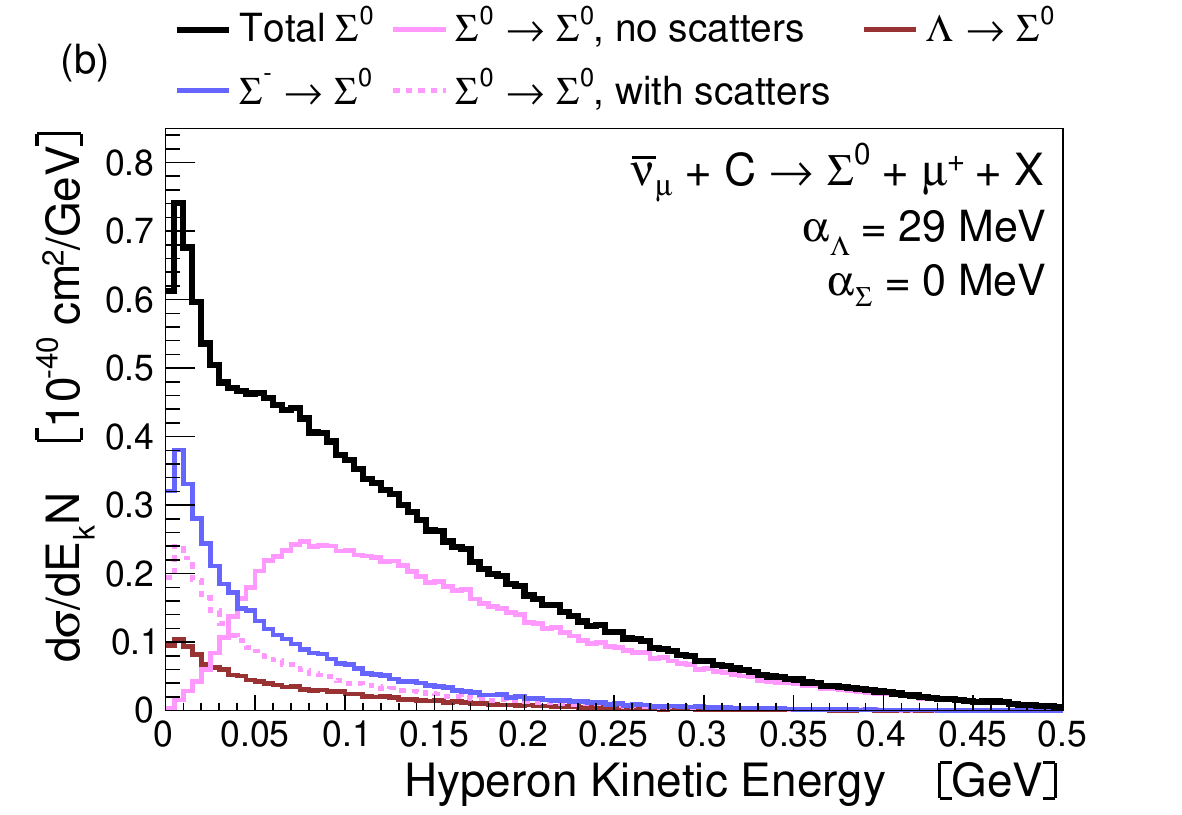}}
\subfigure{\includegraphics[width=0.5\linewidth,height=4.9cm]{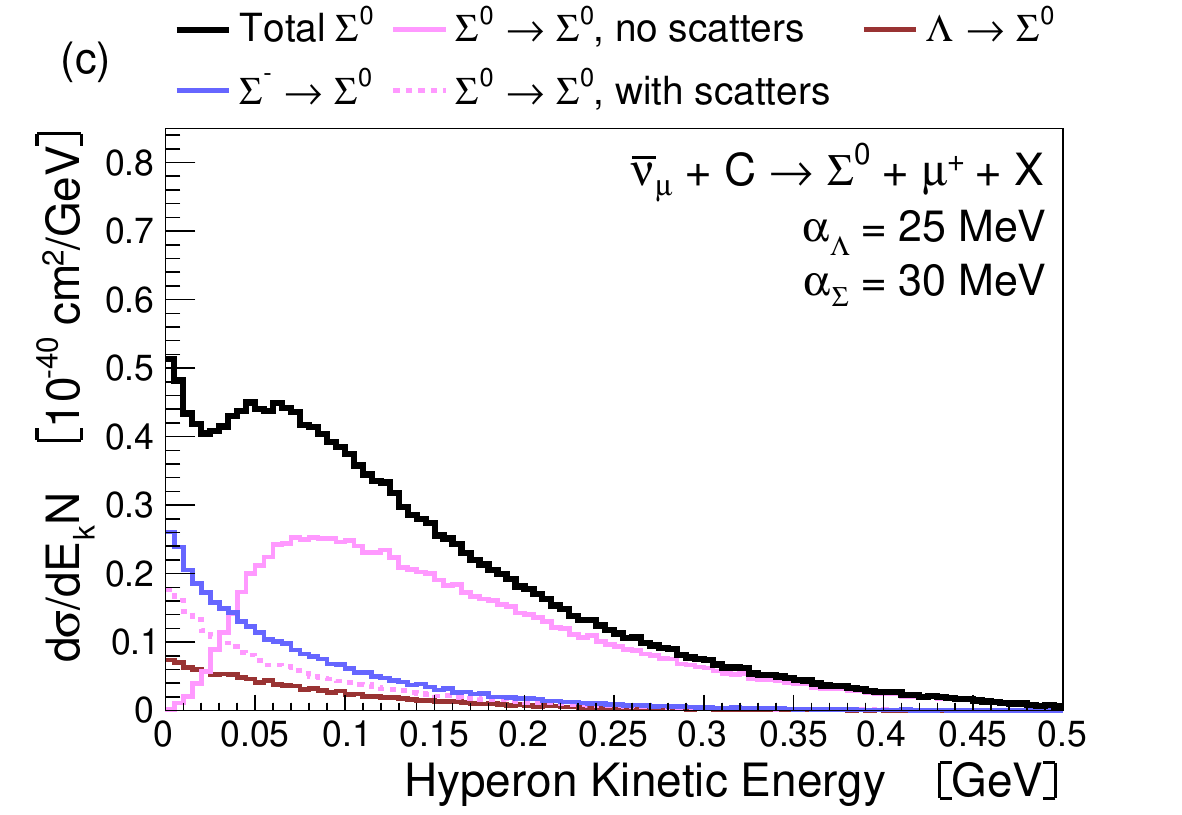}}%
\subfigure{\includegraphics[width=0.5\linewidth,height=4.9cm]{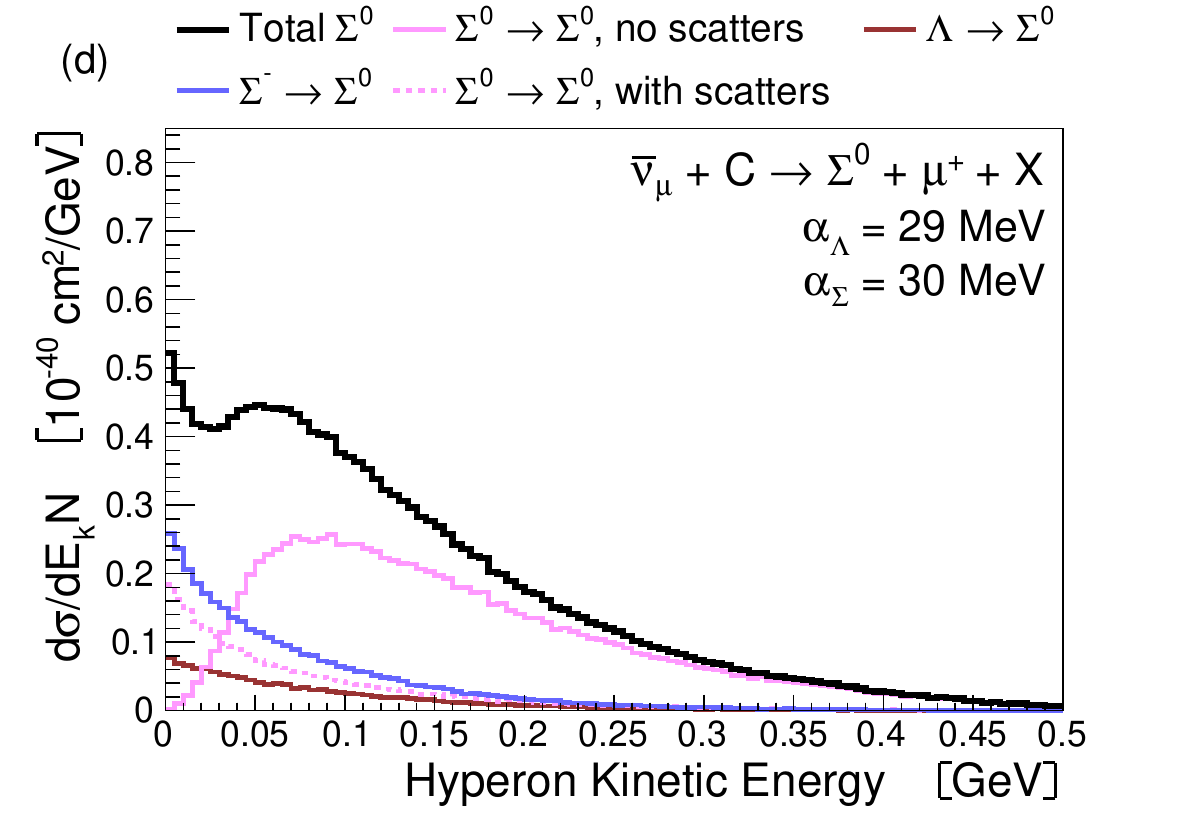}}
\caption{Distribution of $\Sigma^{0}$ kinetic energies after FSI for several choices of the hyperon-nucleus potentials.}
 \label{UnStacked_Sigma_FSI_Rodriguez}
\end{figure*}

\begin{figure*}
\subfigure{\includegraphics[width=0.5\linewidth,height=4.9cm]{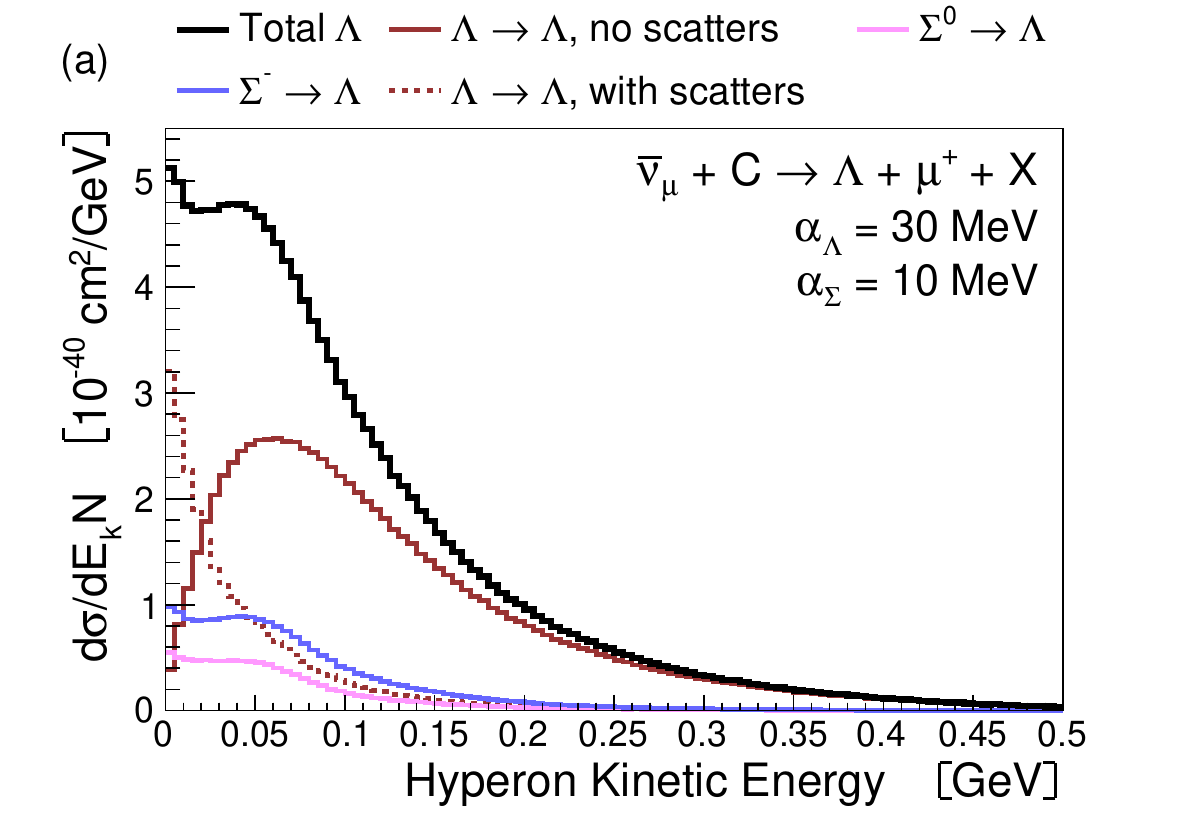}}%
\subfigure{\includegraphics[width=0.5\linewidth,height=4.9cm]{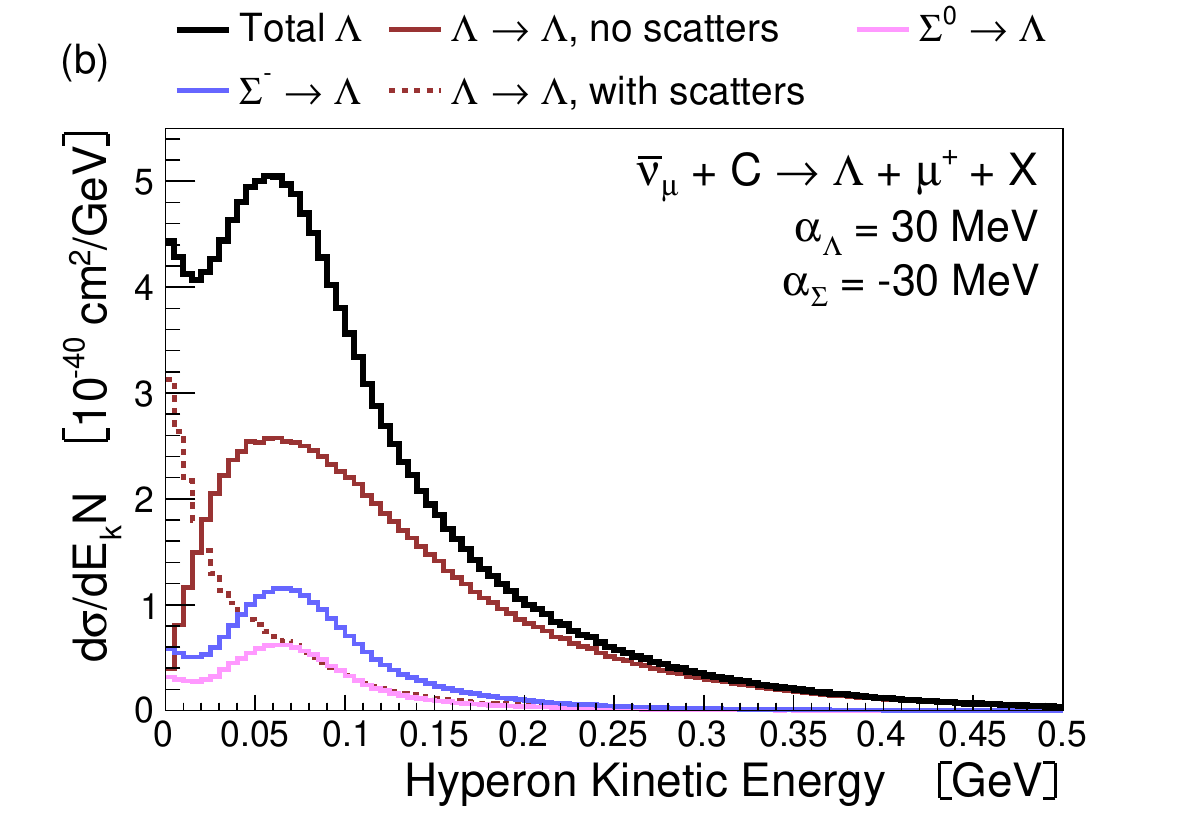}}
\subfigure{\includegraphics[width=0.5\linewidth,height=4.9cm]{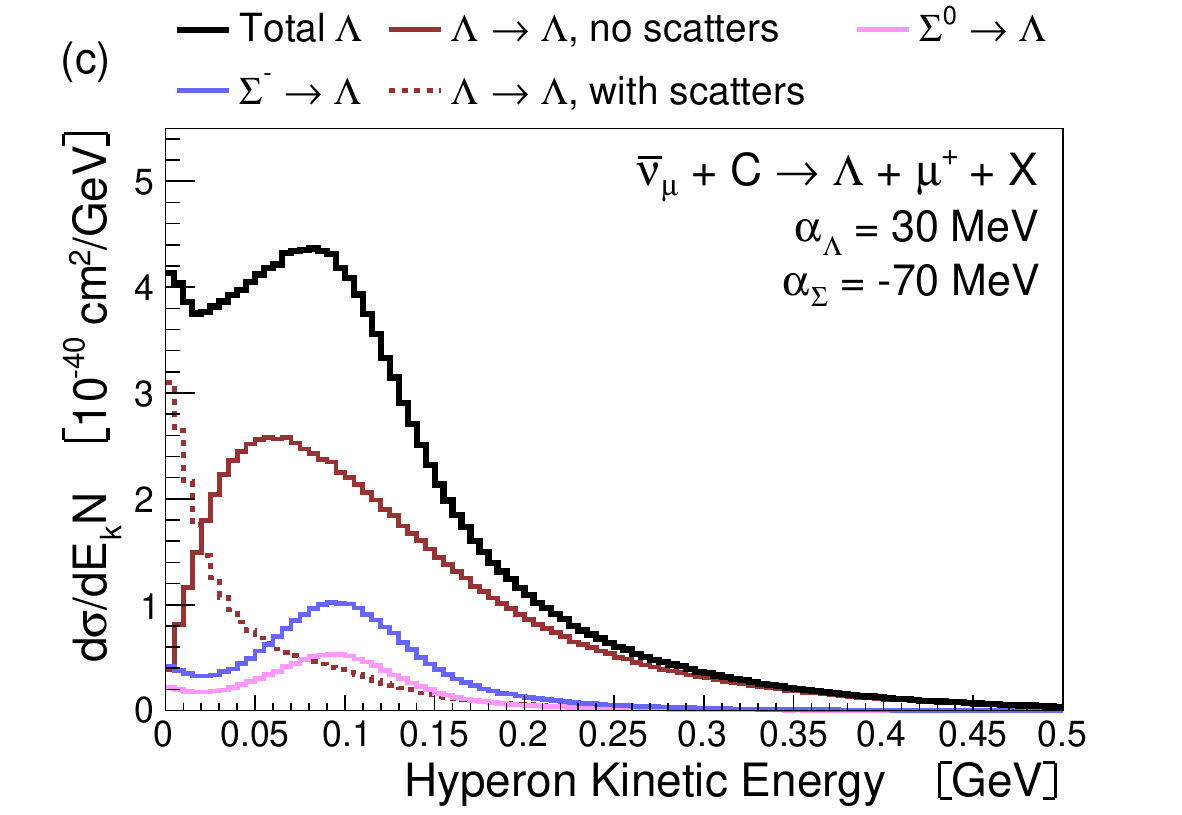}}%
\subfigure{\includegraphics[width=0.5\linewidth,height=4.9cm]{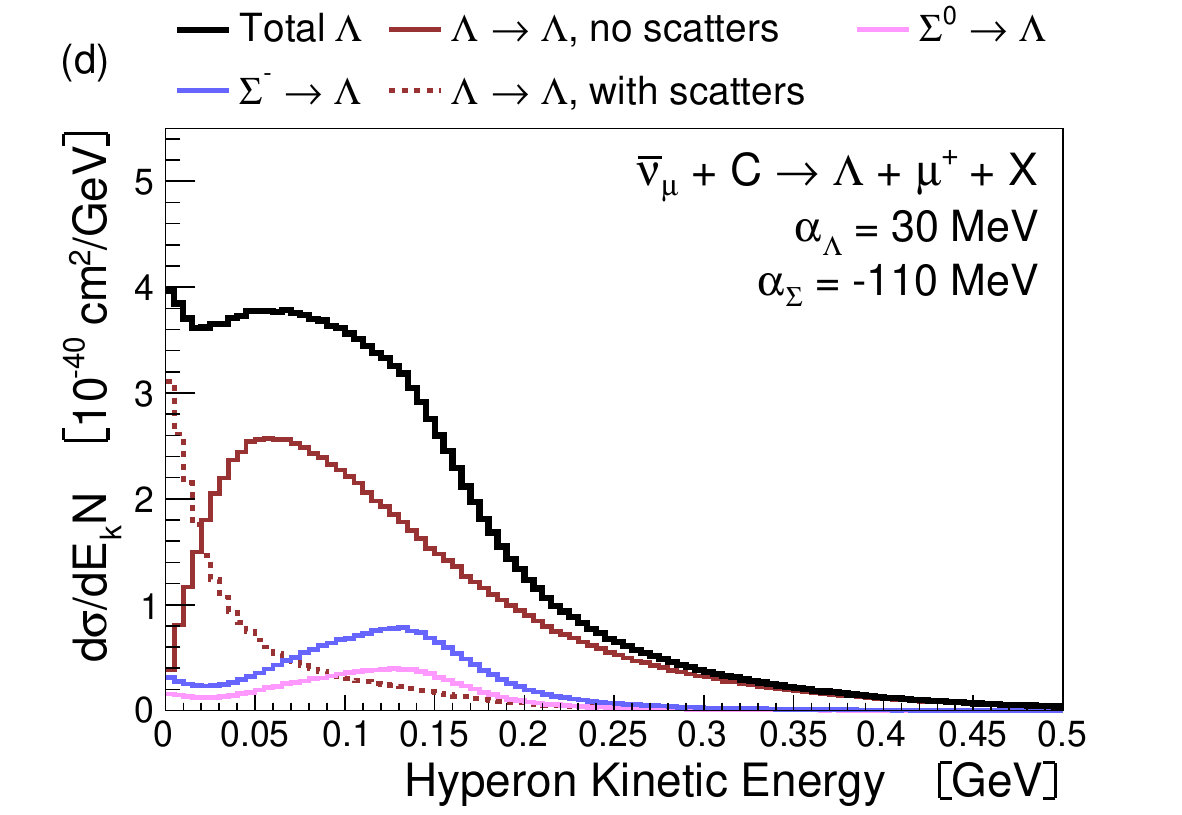}}
\subfigure{\includegraphics[width=0.5\linewidth,height=4.9cm]{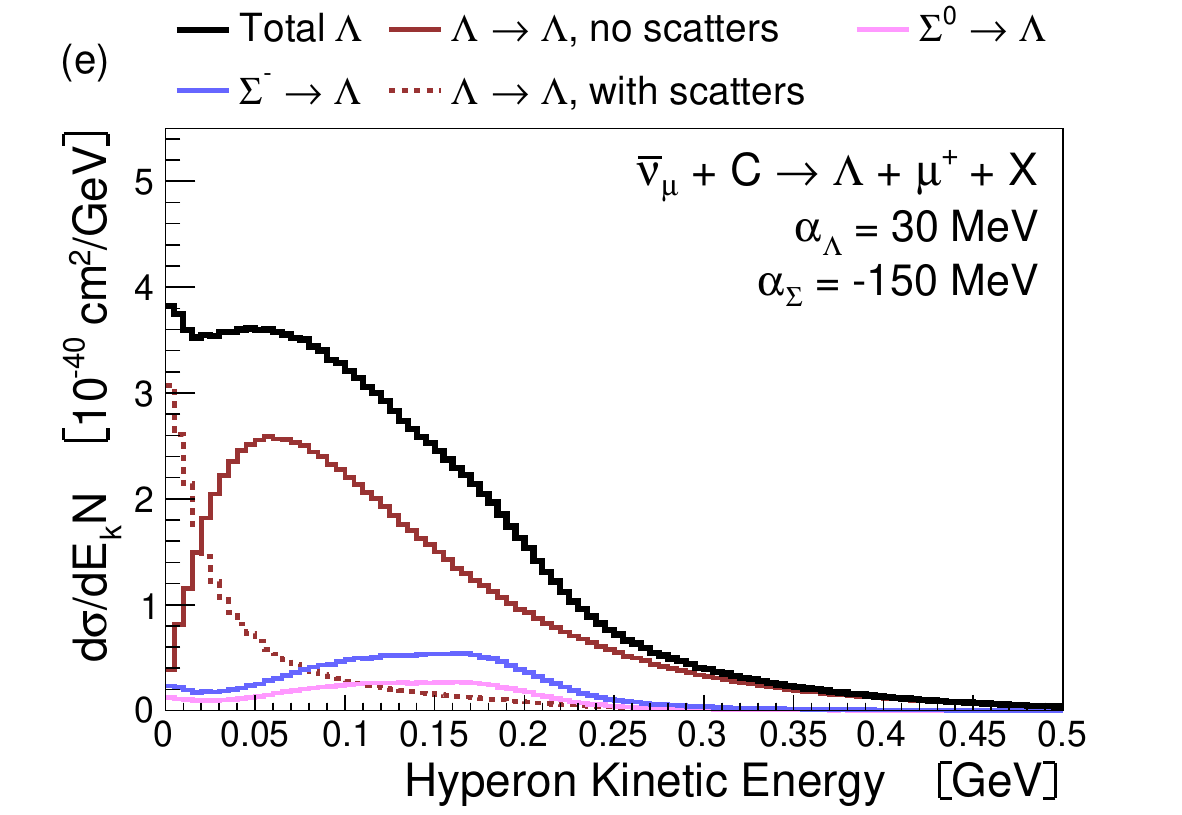}}
\caption{Distribution of $\Lambda$ kinetic energies after final state interactions for several values of the $\Sigma$ nucleus potential. The total differential cross sections are indicated by the solid black line, equal to the sum of the other distributions.}
\label{UnStacked_Lambda_FSI_Saha}
\end{figure*}

\begin{figure*}
\subfigure{\includegraphics[width=0.5\linewidth,height=4.9cm]{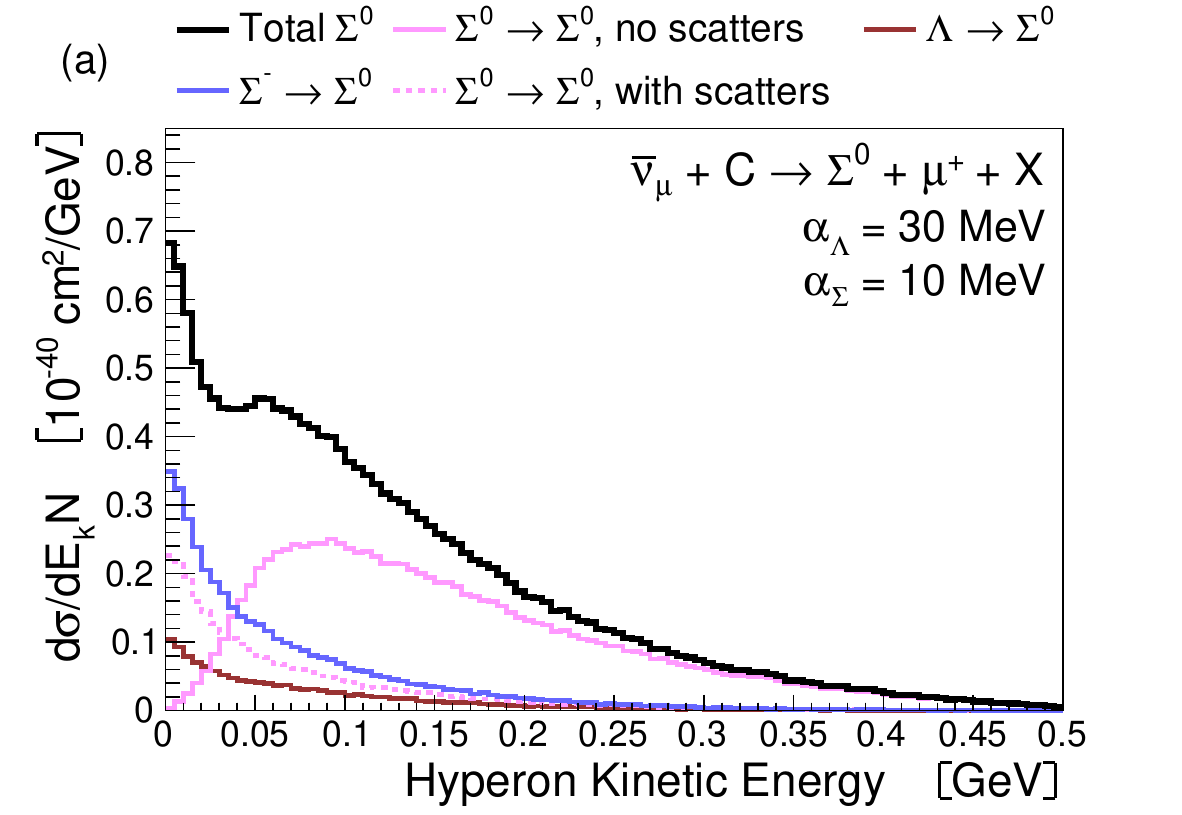}}%
\subfigure{\includegraphics[width=0.5\linewidth,height=4.9cm]{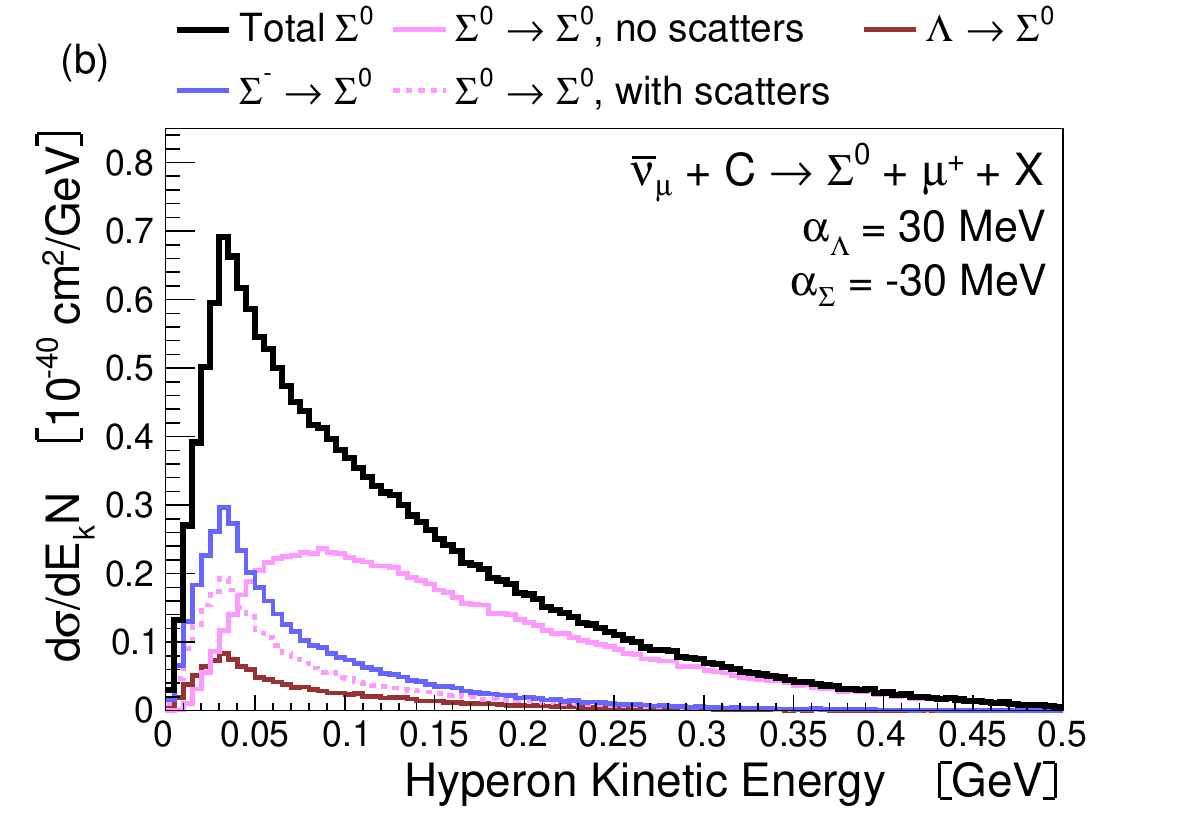}}
\subfigure{\includegraphics[width=0.5\linewidth,height=4.9cm]{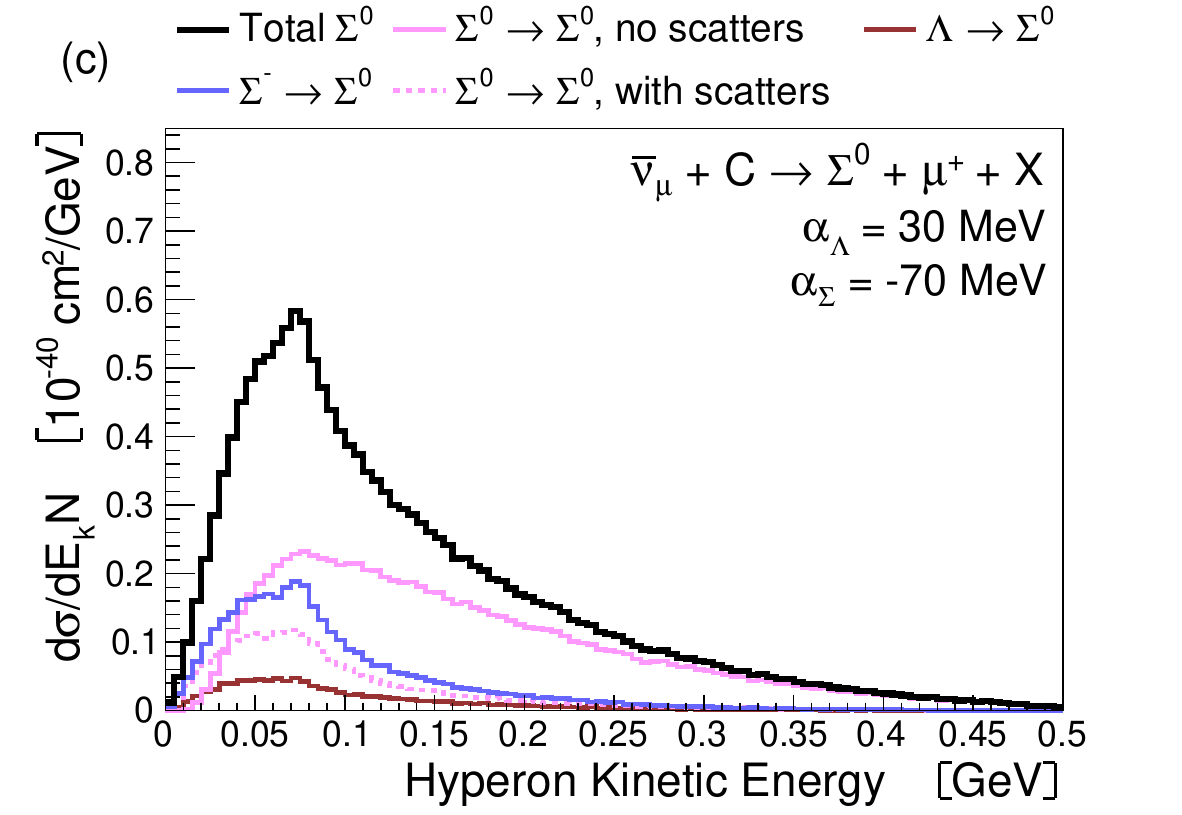}}%
\subfigure{\includegraphics[width=0.5\linewidth,height=4.9cm]{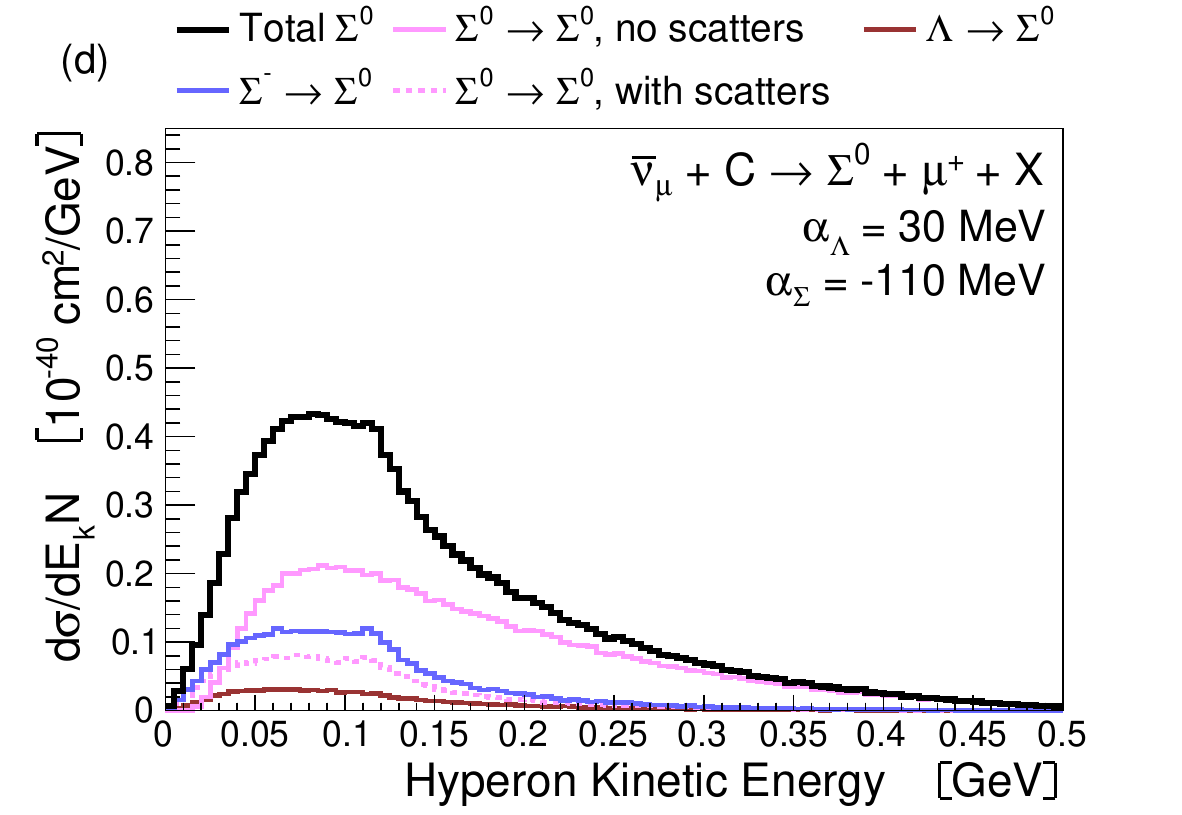}}
\subfigure{\includegraphics[width=0.5\linewidth,height=4.9cm]{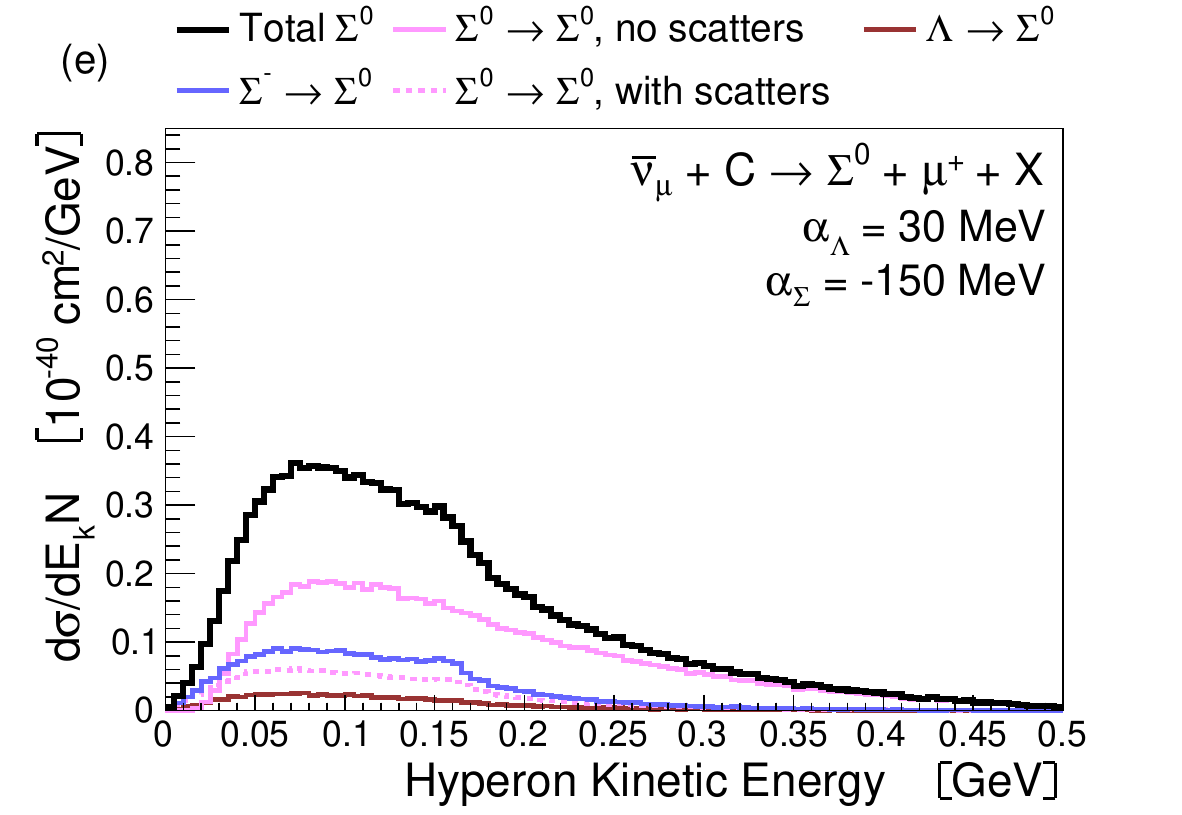}}
\caption{Distribution of $\Sigma$ kinetic energies after final state interactions for several values of the $\Sigma$ nucleus potential.The total differential cross sections are indicated by the solid black line, equal to the sum of the other distributions.}
\label{UnStacked_Sigma_FSI_Saha}
\end{figure*}

To study FSI and the response to changing these potentials, figures \ref{UnStacked_Lambda_FSI_Rodriguez} - \ref{UnStacked_Sigma_FSI_Saha} show the hyperon kinetic energy distribution broken down into different contributions, such as $\Lambda$'s propagating through the nucleus with no reinteraction, $\Lambda$'s that underwent intranuclear rescattering but survived, or $\Sigma^0 \to \Lambda$ conversions. The total differential cross section is shown as a solid black curve.

Figures \ref{UnStacked_Lambda_FSI_Rodriguez} and \ref{UnStacked_Sigma_FSI_Rodriguez} show the effect of varying the $\Lambda$ nucleus potential within the range allowed in~\cite{Rodriguez-Sanchez:2018oqv}. Figure \ref{UnStacked_Lambda_FSI_Rodriguez} shows a very small change in the $\Lambda$ cross section is observed at low energy (below the level of the potential), corresponding to approximately a 10\% decrease at zero kinetic energy and only changes the total cross section by approximately 1\%. The change to the $\Sigma^0$ production cross section is negligible. The effect of changing the $\Sigma$ potential is far more significant. 

Figures \ref{UnStacked_Lambda_FSI_Saha} and \ref{UnStacked_Sigma_FSI_Saha} display the hyperon kinetic energy spectrum for five values of the $\Sigma$ potential, keeping the $\Lambda$ potential fixed with $\alpha=30$ MeV, similar to the study presented in~\cite{Saha:2004ha}. The effect of changing the $\Sigma$ nucleus potential is most apparent in the $\Sigma^{-}\to\Lambda$ and $\Sigma^{0}\to\Lambda$ contributions, which translate to higher kinetic energies and the potential becomes more strongly repulsive. Mostly the distribution for hyperons of kinetic energy above approximately 0.25 GeV is unchanged, but the modification to the distribution below this level is significant. There is potential for neutrino experiments to use these features to constrain this potential if they are sensitive to hyperons in this range of energies.

\newpage

\section{Realistic Fluxes}
\label{Section6}

\subsection{Choice of Experiments}

Real experiments use fluxes with a mixture of neutrino energies and flavours on a variety of nuclear targets. We produce predictions for experiments by convolution of differential cross sections of the form shown above with neutrino energy distributions. We present predictions for the experiment-like setups shown in table \ref{Experiments}.

\begin{table}[H]
    \centering
    \begin{tabular}{|c|c|c|}
    \hline
      Beam   & Target  & Experiment(s) \\
      \hline
       Booster (BNB)  & Ar & SBND, MicroBooNE \\
       NuMI & C & NOvA \\
       DUNE Flux & Ar & DUNE \\
       NOMAD Flux & C & NOMAD \\
       NuSTORM & C,Ar & NuSTORM \\
       \hline
    \end{tabular}
    \caption{List of simulated fluxes, targets and corresponding experiments. Proposals to combine NuSTORM flux with several detector designs have been made, including various scintillating targets and argon TPCs. With the exception of NOMAD, these beams can operate in neutrino mode (forward horn current, FHC) and antineutrino mode (reverse horn current, RHC) where the flux is dominated by neutrinos/antineutrinos respectively.}
    \label{Experiments}
\end{table}

 We study nuclear effects and variations of axial mass on several variables using these realistic fluxes and targets: The $Q^2$ distributions, the hyperon momentum and lepton kinetic energy, the angle of the charged lepton with respect to the beam axis, $\theta_{\mu}$ and the opening angle between the hyperon and lepton. We will concentrate on the dominant channel, $\Lambda$ production.

\subsection{Final State Interactions, Axial Mass and Different Observables}

\begin{figure}[H]
    \centering
    \includegraphics[width=\linewidth,height=5.4cm]{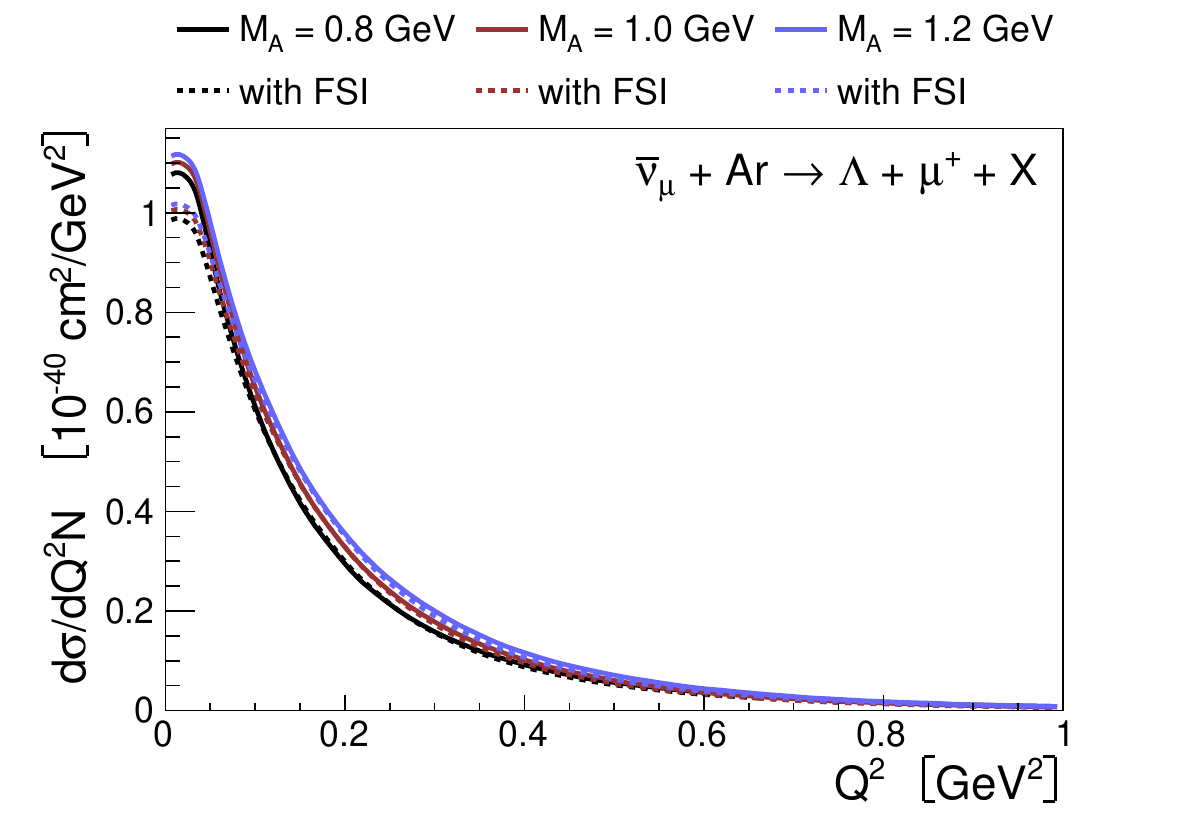}
    \caption{Differential cross section with respect to $Q^2$ using the BNB neutrino mode flux on argon.}
    \label{BNB_FHC_MA_Lambda_Q2}
\end{figure}

\begin{figure}[H]
    \centering
    \includegraphics[width=\linewidth,height=5.4cm]{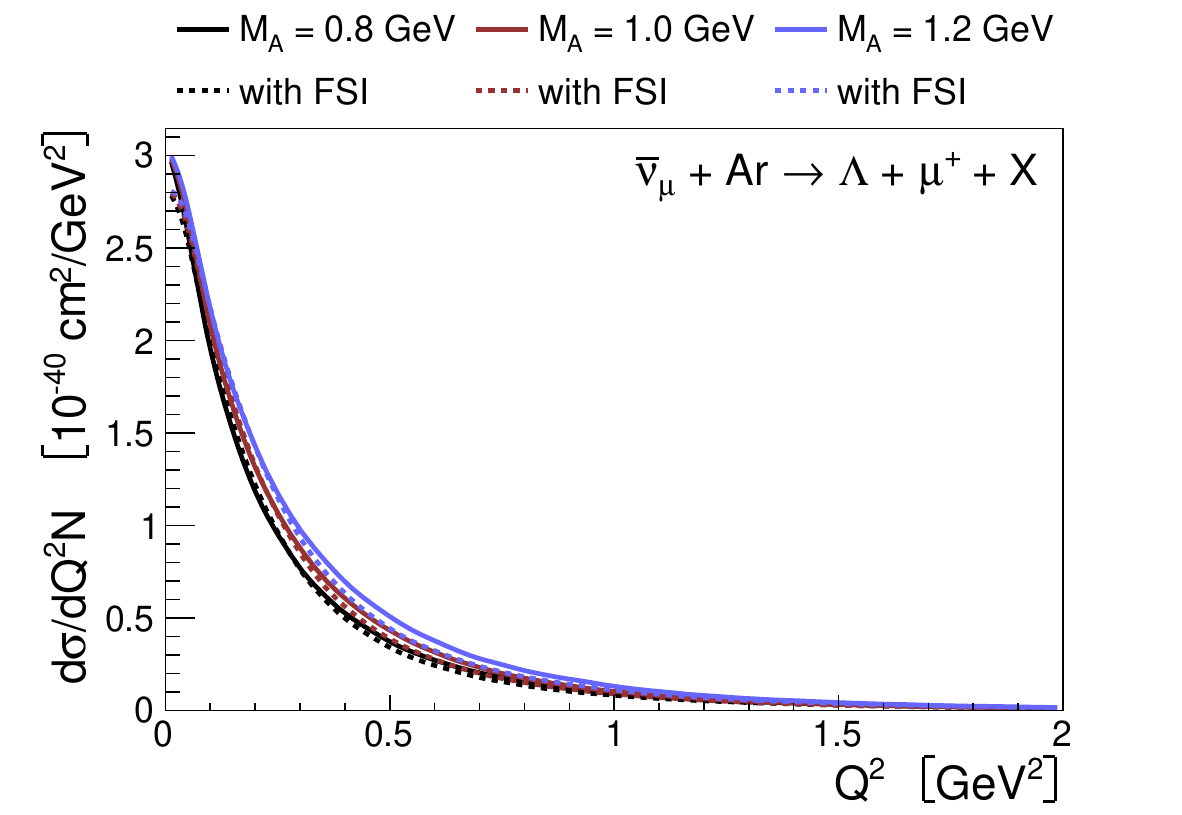}
    \caption{Differential cross section per nucleon with respect to $Q^2$ using the DUNE flux on argon.}
    \label{DUNE_FHC_MA_Lambda_Q2} 
\end{figure}

The $Q^2$ distributions for the BNB and DUNE in neutrino mode are presented in figures \ref{BNB_FHC_MA_Lambda_Q2} and \ref{DUNE_FHC_MA_Lambda_Q2}, two LAr experiments using different fluxes. The DUNE prediction is significantly larger resulting from higher average antineutrino energies. We see this variable fairly insensitive to the axial mass, even using the more energetic DUNE flux. The main result of the varied axial masses is a small enhancement of the cross section in the mid $Q^2$ region. 

\begin{figure}[H]
    \centering
    \includegraphics[width=\linewidth,height=5.4cm]{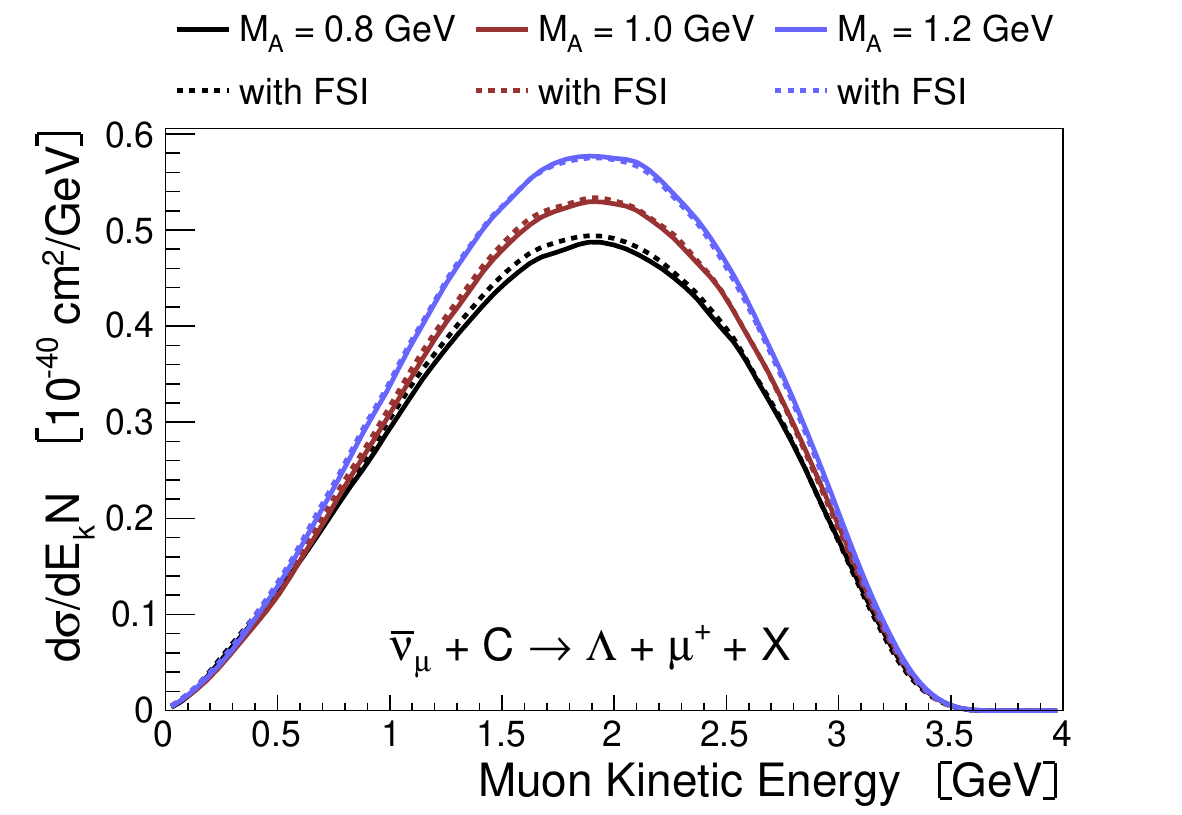}
    \caption{Differential cross section with respect to the muon kinetic energy using the NuSTORM flux on carbon.}
    \label{NuSTORM_MA_Lambda_Lepton_KE}
\end{figure}

\begin{figure}[H]
    \centering
    \includegraphics[width=\linewidth,height=5.4cm]{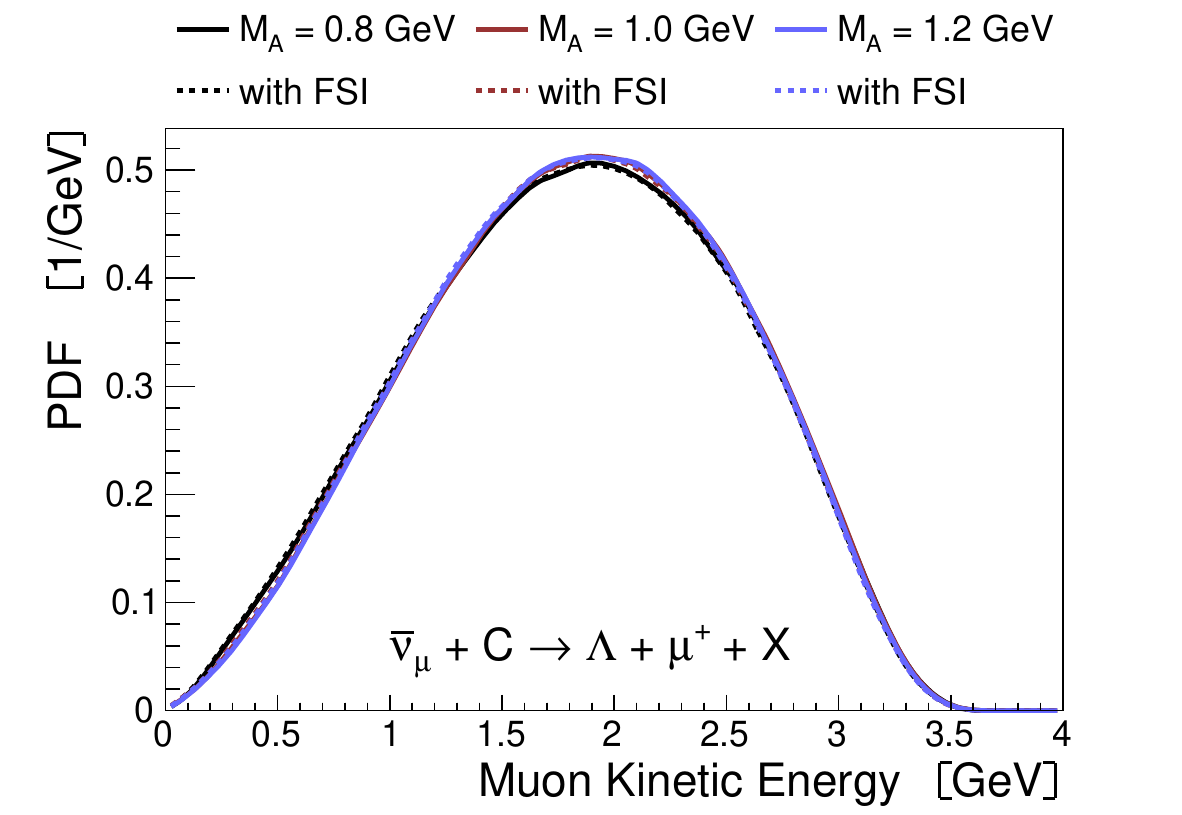}
    \caption{Shape of the differential cross section with respect to the muon kinetic energy using the NuSTORM flux on carbon.}
    \label{ShapeOnly_NuSTORM_MA_Lambda_Lepton_KE}
\end{figure}

Typically, the lepton will receive a large fraction of the energy budget of the process due to the low $Q^2$ dominance shown in figures 6-10 and the lack of dissipation due to nuclear effects. As a result the lepton will typically travel furthest in a detector, depending on its kinetic energy. Figure \ref{NuSTORM_MA_Lambda_Lepton_KE} shows the differential cross section with respect to $\mu^+$ kinetic energies using the NuSTORM flux on an argon target. A significant change results from varying the axial mass, however figure \ref{NuSTORM_MA_Lambda_Lepton_KE} suggests less sensitivity in the shape. To study the shape effects more closely, we redraw the same distributions, all normalised to 1 in figure \ref{ShapeOnly_NuSTORM_MA_Lambda_Lepton_KE} displaying a small change in the position of the modal kinetic energy.

\begin{figure}[H]
    \centering
    \includegraphics[width=\linewidth,height=5.4cm]{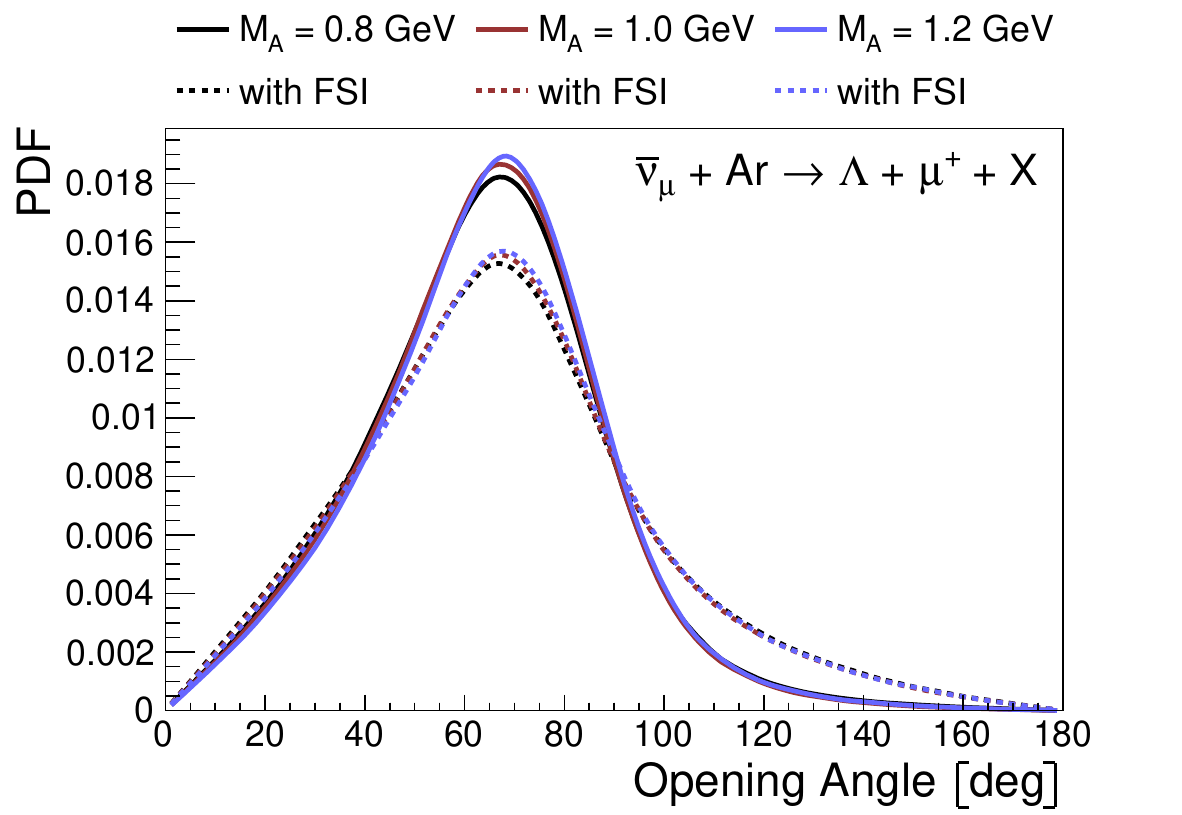}
    \caption{Distribution of opening angles between the $\mu^+$ and $\Lambda$ produced using the BNB neutrino mode flux on argon.}
    \label{BNB_FHC_MA_Lambda_Opening_Angle}
\end{figure}

\begin{figure}[H]
    \centering
    \includegraphics[width=\linewidth,height=5.4cm]{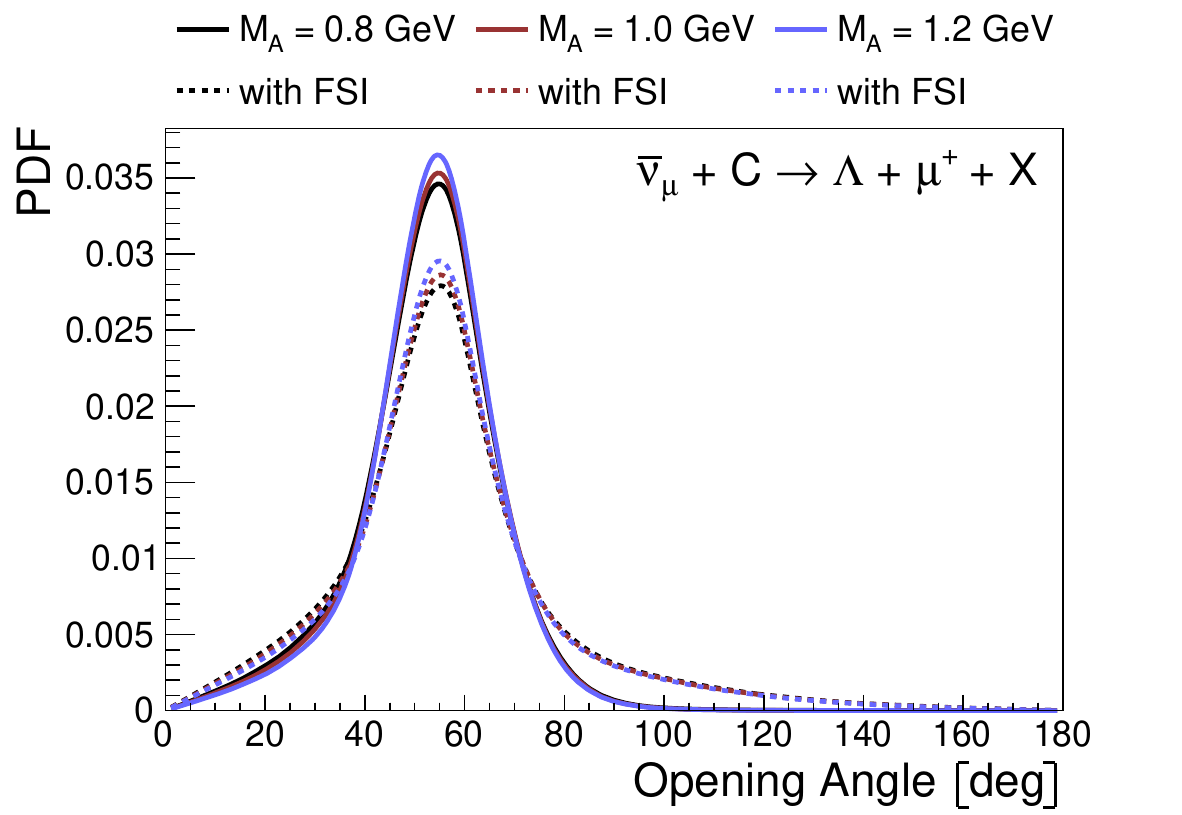}
    \caption{Distribution of opening angles between the $\mu^+$ and $\Lambda$ produced using the NOMAD flux on carbon.}
    \label{NOMAD_MA_Lambda_Opening_Angle}
\end{figure}

The distribution of opening angles between the hyperon and muon in $\Lambda$ events using the BNB neutrino mode flux and the NOMAD flux are presented in figures \ref{BNB_FHC_MA_Lambda_Opening_Angle} and \ref{NOMAD_MA_Lambda_Opening_Angle}. The effect of final state interactions is seen through a broadening of the distribution; there is a significantly greater chance of the opening angle exceeding 90 degrees. The effect of varying the axial mass is small, producing a very slight change in the probability of observing a smaller opening angle, below approximately 50 degrees. 

\begin{figure}[H]
    \centering
    \includegraphics[width=\linewidth,height=5.4cm]{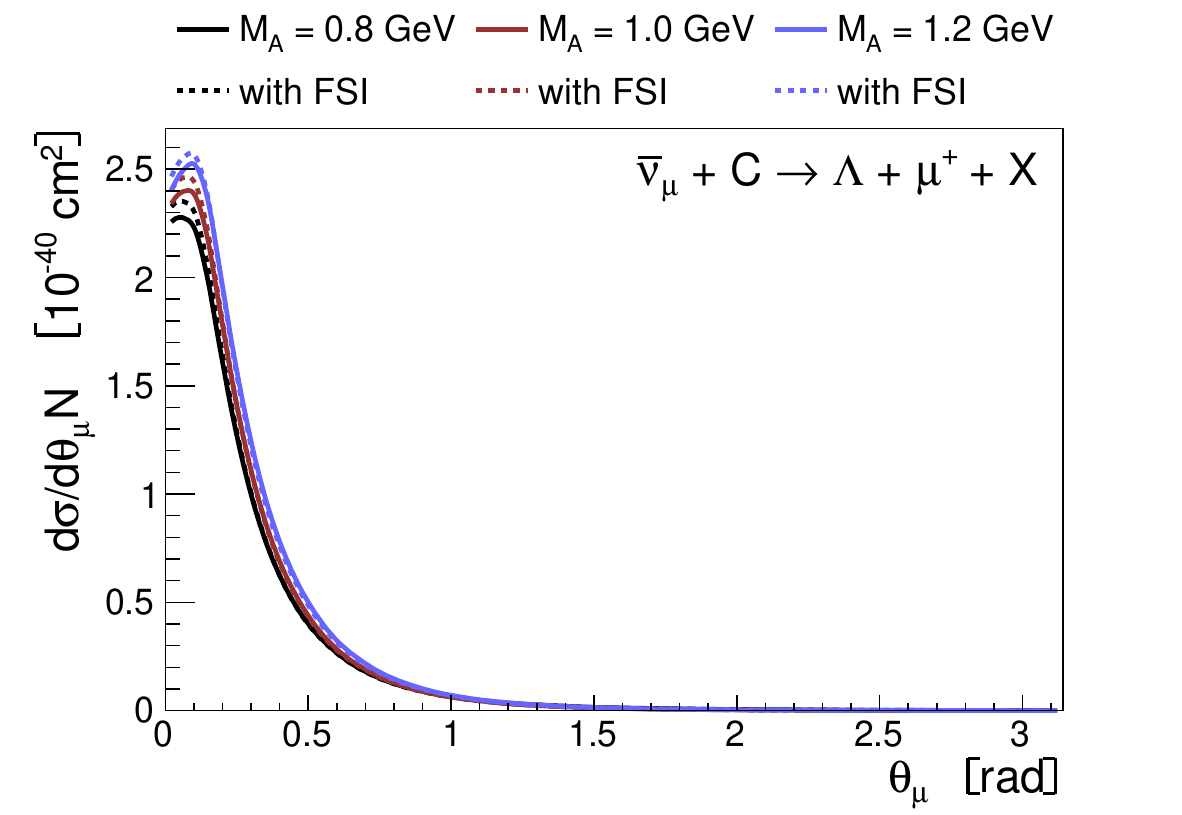}
    \caption{Differential cross section with respect to lepton scattering angle using the NOvA neutrino mode flux on carbon.}
    \label{NOvA_FHC_MA_Lambda_Lepton_Angle}
\end{figure}

\begin{figure}[H]
    \centering
    \includegraphics[width=\linewidth,height=5.4cm]{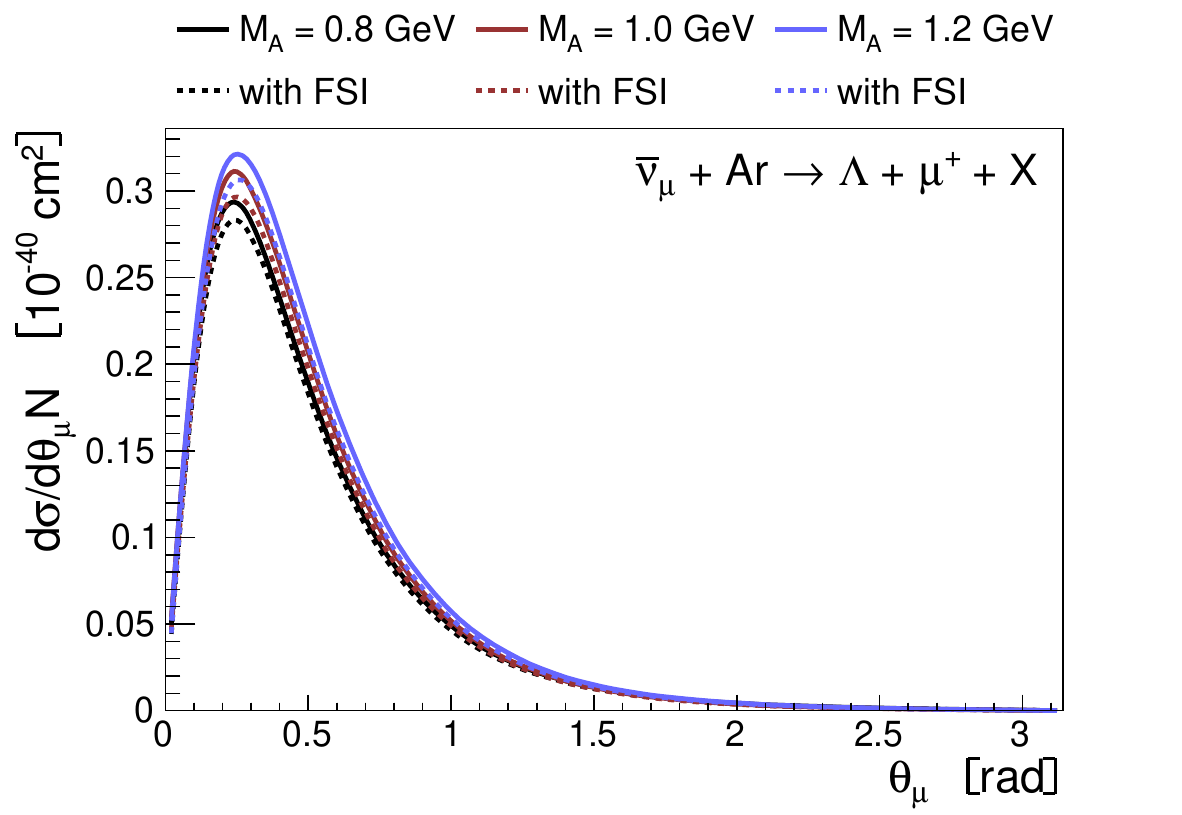}
    \caption{Differential cross section with respect to lepton scattering angle using the BNB neutrino mode flux on argon.}
    \label{BNB_FHC_MA_Lambda_Lepton_Angle}
\end{figure}

A related variable is the scattering angle of the $\mu^+$, shown for the NOvA flux in figure \ref{NOvA_FHC_MA_Lambda_Lepton_Angle} and the BNB flux in figure \ref{BNB_FHC_MA_Lambda_Lepton_Angle}. The more energetic flux from NOvA preferentially produces leptons at very small angles in comparison to the BNB, the result of the sharpening of the $Q^2$ curve as the neutrino energy increases.

\begin{figure}[H]
    \centering
    \includegraphics[width=\linewidth,height=5.4cm]{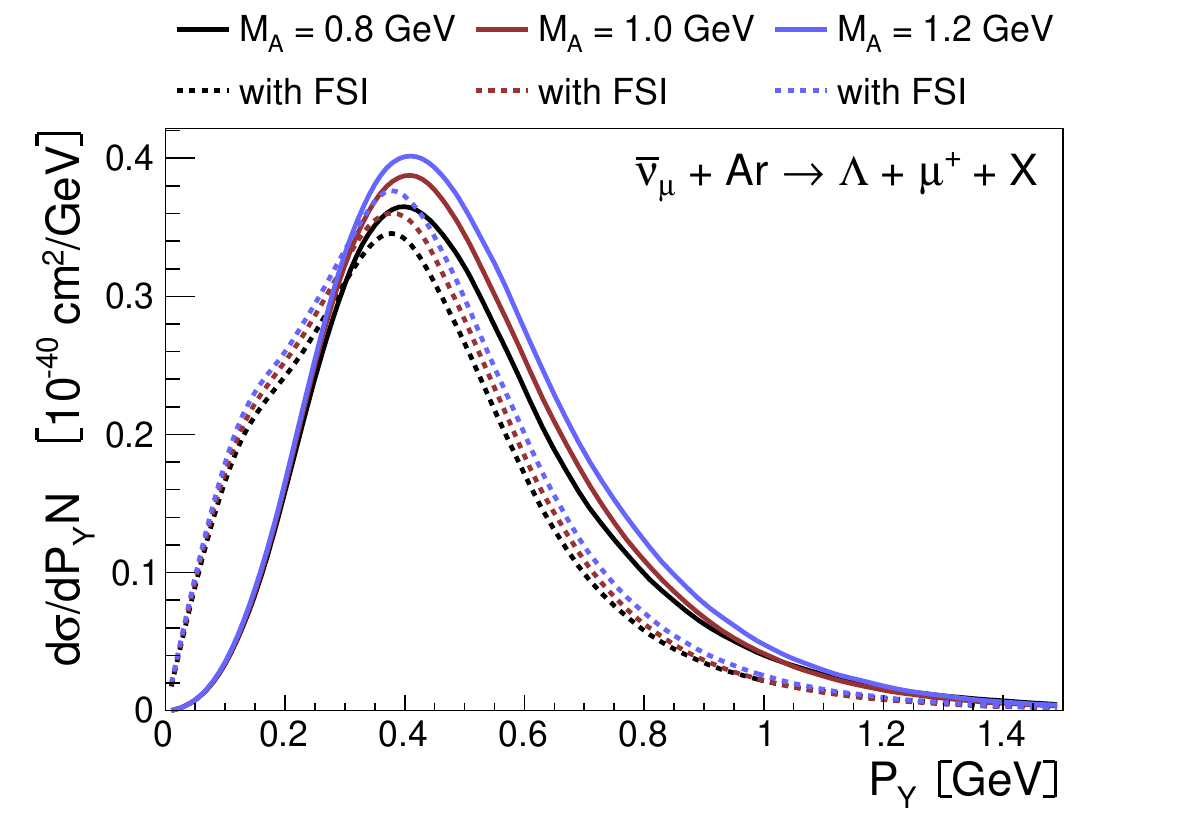}
    %\caption{Differential cross section for $\bar{\nu}_{\mu} + \textrm{Ar} \to \Lambda + \mu^+ + X$ with respect to the $\Lambda$ momentum using the BNB neutrino mode flux.}
    \caption{Differential cross section with respect to the $\Lambda$ momentum produced using the BNB neutrino mode flux.}
    \label{BNB_FHC_MA_Lambda_Hyperon_Mom}
\end{figure}

\begin{figure}[H]
    \centering
    \includegraphics[width=\linewidth,height=5.4cm]{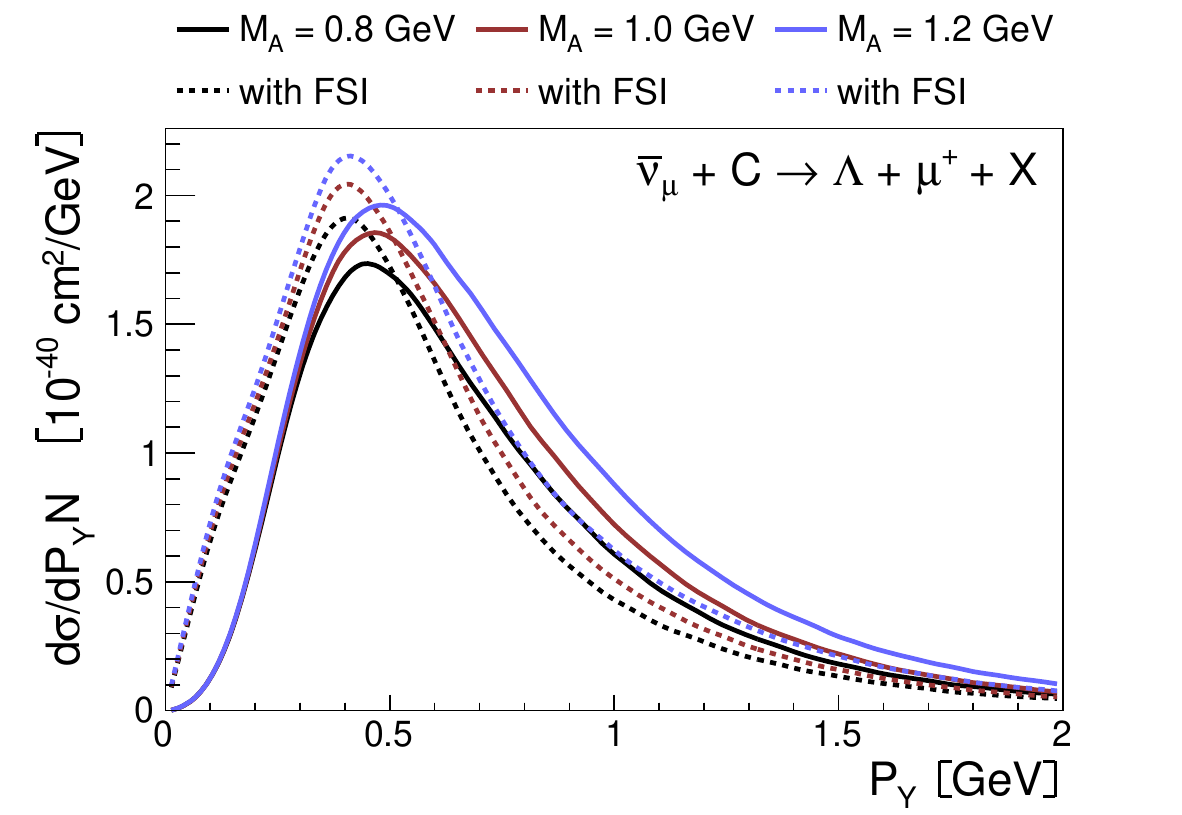}
    \caption{Differential cross section with respect to the $\Lambda$ momentum produced using the NOMAD flux.}
    \label{NOMAD_MA_Lambda_Hyperon_Mom}
\end{figure}

An important feature in identifying events of this type from background is the displacement between the primary vertex and the hyperon decay in relation to the spatial resolution of the detector. The decay length $\lambda$ of the hyperon depends on the lifetime $\tau$ and the 3 momentum $|\textbf{p}|$ through $\lambda = \tau |\textbf{p}|/M_Y$. Figures \ref{BNB_FHC_MA_Lambda_Hyperon_Mom} and \ref{NOMAD_MA_Lambda_Hyperon_Mom} show the distribution of $\Lambda$ momenta produced using the BNB FHC and NOMAD fluxes, the least energetic and most energetic fluxes studied. Comparison of these distributions shows the maximum hyperon momentum only increases by a third whilst the typical neutrino energy has been changed by over an order of magnitude. The position of the peak of the distribution does not change either. Final state interactions have the effect of shifting this peak to lower values, which can be understood as the hyperon dissipating kinetic energy in the nucleus as it escapes. The shape of this distribution is sensitive to FSI, with little enhancement seen below the peak of this distribution, however there is an increase beyond it, most noticeably in the NOMAD case.

\begin{figure}[H]
    \centering
    \includegraphics[width=\linewidth,height=5.4cm]{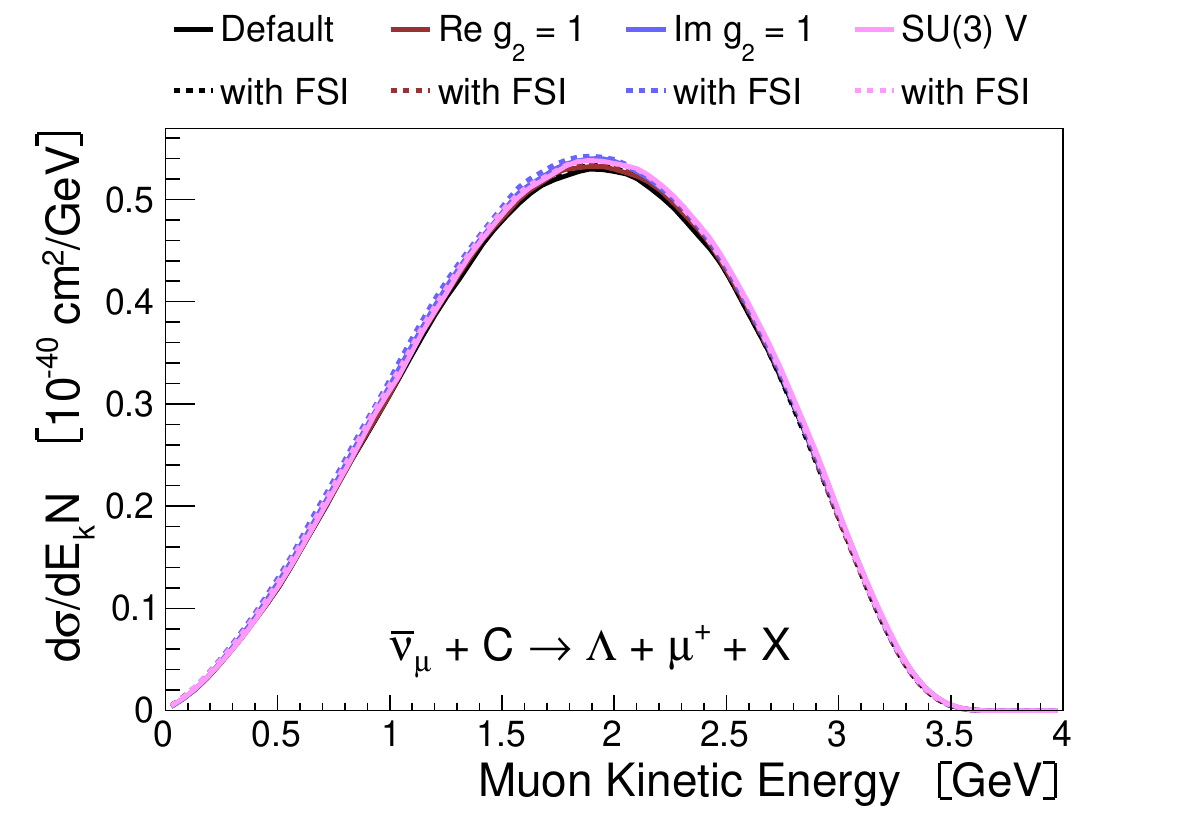}
    \caption{Differential cross section for $\bar{\nu}_{\mu} + \textrm{C} \to \Lambda + \mu^+ + X$ using the NuSTORM flux, including second class current and SU(3) symmetry breaking effects.}
    \label{NuSTORM_BSM_Lambda_Lepton_KE}
\end{figure}

The NuSTORM beam differs from existing neutrino beams in that neutrinos are primarily generated from decaying muons stored in an accelerator, with the potential to reduce neutrino flux uncertainties to $\lesssim$ 1\%\cite{Adey:2017dvr}, beneficial for the study of subtle effects such as SCC. We show the effect of second class current and SU(3) symmetry breaking on the lepton kinetic energy distribution in figure \ref{NuSTORM_BSM_Lambda_Lepton_KE}. There is a very small increase in the cross section, on the order of a few percent, which persists after final state interactions.

% Q^2 plots

%\vspace{0.0cm}

% Lepton KE plots

% Opening angle

% Hyperon momentum

\section{Conclusions}
\label{Section7}

We have studied the properties of hyperon production in neutrino experiments, including the effects of variations in the axial mass, SU(3) symmetry breaking, and time reversal respecting and violating second class currents. We find the total cross sections for hyperon production from free nucleons is sensitive to variations in axial mass, with a relative change that increases with neutrino energy and is largest for $\Lambda$ production. The SU(3) symmetry breaking corrections have the effect of increasing the $\Lambda$ production rate and decreasing the $\Sigma^0$ rate, suggesting a comparison of these cross sections may offer better sensitivity than separate measurements.

A nuclear cascade is used to propagate the hyperon through the rest of the nucleus, resulting in significant changes in contributions of individual channels to the total hyperon production cross section. The $\Lambda$ production rate receives a significant enhancement from conversion of $\Sigma$ to $\Lambda$. The SU(3) quark model predicts a ratio between the $\Sigma^-$ and $\Sigma^0$ production rates of 1/2, which is approximately maintained after final state interactions. A small number of $\Sigma^+$ are produced through charge exchange in hyperon-nucleon interactions. Hyperons are subject to a simple test potential which often results in their absorption into the nucleus leading to a suppression of low $Q^2$ interactions. 

We perform cross section calculations for experiments using realistic neutrino energy distributions and suitable nuclear targets, and study the effect of using different fluxes and axial masses on observable quantities. Changes in axial mass are most visible in the differential cross section with respect to the kinetic energy of the outgoing lepton and momentum of the outgoing hyperon, however the shape of the former distribution is not sensitive to this parameter. Larger axial mass values result in an enhancement of the production rate of high momentum hyperons. 

\vspace{0.7cm}

\newpage

\begin{acknowledgments}
This research was funded in part by Science and Technology Facilities Council grant number ST/S000895/1, by  Polish Ministry of Science and Higher Education Grant DIR/WK/2017/05 and also by NCN Opus Grant 2016/21/B/ST2/01092.
\end{acknowledgments}


%% Convert this to BIBTEX when ready

%\vspace{2cm}

\bibliography{HyperonProduction}

\begin{comment}

\end{comment}

\end{document}